\documentclass[aip, amsmath, amssymb, preprint]{revtex4-1}

\usepackage{graphicx}
\usepackage{dcolumn}
\usepackage{bm}
\usepackage{caption}
\usepackage{subcaption} 
\usepackage[colorlinks=true,
            linkcolor=blue,
            urlcolor=blue,
            citecolor=blue]{hyperref}
\usepackage{calrsfs}
\DeclareMathAlphabet{\pazocal}{OMS}{zplm}{m}{n} 

\newcommand{\Lb}{\pazocal{L}}

\usepackage{color,colortbl}
\definecolor{Gray}{gray}{0.9}

\usepackage[color=yellow!50,colorinlistoftodos,prependcaption, textsize=small]{todonotes}
\usepackage{color}
\usepackage{tikz}
\usetikzlibrary{shapes}
\newcommand{\markerDone}{\raisebox{0pt}{\tikz{\node[draw,scale=0.6,regular polygon, regular polygon sides=4,fill=black!20!black](){};}}}
\newcommand{\markerDtwo}{\raisebox{0pt}{\tikz{\node[draw,scale=0.6,circle,fill=black!20!black](){};}}}
\newcommand{\markerDthree}{\raisebox{0pt}{\tikz{\node[draw,scale=0.5,diamond,fill=black!20!black](){};}}}
\newcommand{\markerDfour}{\raisebox{0pt}{\tikz{\node[draw,scale=0.4,regular polygon, regular polygon sides=3,fill=black!20!black,rotate=180](){};}}}
\newcommand{\markerDfive}{\raisebox{0pt}{\tikz{\node[draw,scale=0.4,regular polygon, regular polygon sides=3,fill=black!20!black,rotate=0](){};}}}
\usepackage{xcolor}
\usepackage[export]{adjustbox}

\makeatletter
\def\@email#1#2{%
 \endgroup
 \patchcmd{\titleblock@produce}
  {\frontmatter@RRAPformat}
  {\frontmatter@RRAPformat{\produce@RRAP{*#1\href{mailto:#2}{#2}}}\frontmatter@RRAPformat}
  {}{}
}%
\makeatother

\begin{document}

\title{\textbf{Probing aqueous interfaces with spin defects}}

\author{Alfonso Castillo}
\affiliation{Department of Physics, The University of Chicago, Chicago, Illinois 60637, USA}

\author{Gustavo R. P\'erez-Lemus}
\affiliation{Pritzker School of Molecular Engineering, The University of Chicago, Chicago, Illinois 60637, USA}
\affiliation{Department of Chemical and Biomolecular Engineering, Tandon School of Engineering, New York University, New York, New York 10012, USA}

\author{Mykyta Onizhuk}
\affiliation{Pritzker School of Molecular Engineering, The University of Chicago, Chicago, Illinois 60637, USA}

\author{Giulia Galli}%
\email{gagalli@uchicago.edu}
\affiliation{Pritzker School of Molecular Engineering, The University of Chicago, Chicago, Illinois 60637, USA}
\affiliation{Department of Chemistry, The University of Chicago, Chicago, Illinois 60637, USA}
\affiliation{Materials Science Division and Center for Molecular Engineering, Argonne National Laboratory, Lemont, Illinois 60439, United States}

\date{\today}

\begin{abstract}
Understanding the physical and chemical properties of aqueous interfaces is  important in diverse fields of science, ranging from biology and chemistry to materials science. In spite of crucial progress in surface sensitive spectroscopic techniques over the past decades, the microscopic properties of aqueous interfaces remain difficult to  measure. 
Here we explore the use of noise spectroscopy to characterize interfacial properties, specifically of quantum sensors hosted in two-dimensional materials in contact with water. We combine molecular dynamics simulations of water/graphene interfaces and the calculations of the spin dynamics of an NV-like color center, and we investigate  the impact of interfacial water and simple ions  on the decoherence time of the defect. We show that the Hahn echo coherence time of the NV center is sensitive to motional narrowing and to the hydrogen bonding arrangement and the dynamical properties of water and ions at the interface. We present results as a function of the liquid temperature, strength of the water-surface interaction,  and for varied mono-valent and di-valent ions, highlighting the broad applicability of near-surface qubits to gain insight into the properties of aqueous interfaces.

\end{abstract}

\maketitle



\section{\label{sec:Intro} Introduction}
Understanding interfacial phenomena at the molecular scale is crucial to gain insight into key physical and chemical processes in areas as diverse as catalysis, drug delivery, biosensing and energy technology~\cite{Gonella:2021}. Although surface sensitive  methods such as X-ray photoelectron, Auger and secondary ion mass spectroscopy can provide  chemical analyses of various surfaces and interfaces, it remains challenging to probe experimentally dynamical phenomena and chemical reactions at interfaces~\cite{Salmeron:2018}. Nuclear Magnetic Resonance (NMR) can in principle enable the investigation of interfacial dynamics at the atomic scale, but its traditional implementation has been limited by the need of large volume samples to obtain a  detectable signal~\cite{Fratila:2011}. In the last decade, the development of NMR techniques based on spin defects in semiconductors, specifically NV-centers in diamond (nano-NMR) has paved the way to non-destructive imaging with unprecedented sensitivity,  at  the level of single molecules \cite{Lovchinsky:2016} or even individual spins~\cite{Muller:2014, Sushkov:2014}.

The high specificity of NV centers, and in particular near-surface ones, is naturally accompanied by their sensitivity to magnetic or electronic noise in their close environment. For example, natural diamond, a typical host of NV centers, with 1.1\% of $^{13}$C, can be isotopically purified to obtain an almost nuclear spin-free environment, thus reducing magnetic noise and enhancing the spin defect coherence~\cite{Balasubramanian:2009,Bar_Gill:2013}. However it is challenging to reduce the so called electric noise arising from surface and interfacial defects, e.g. dangling bonds and charge fluctuations. Promising platforms less prone to electrical noise are 2-dimensional materials that in principle  are free of  surface dangling bonds~\cite{Ye:2019}. Nevertheless, the coherence of spin defects in 2D systems is unavoidably affected by the presence of a 3D substrate~\cite{Onizhuk_Waals:2021}, 
and several recent investigations have focused on the search for van der Waals-bonded  materials hosting qubits~\cite{Kanai:2021, Ali:2022, Ali:2023} with suitable substrates.

In the case of nano-NMR at solid-liquid interfaces and, in particular, aqueous interfaces, the diffusion of molecules and ions in the liquid sample~\cite{Kong:2015, Pham:2016, Aslam:2017} may limit the  spectral resolution, thus posing additional challenges relative to nano-NMR at solid surfaces.  Strategies such as liquid confinement~\cite{cohen_confined_2020, Staudenmaier:2022} and the use of highly viscous samples~\cite{Mamin:2013, Muller:2014, Staudacher:2013, Yang:2022} have been proposed to mitigate diffusion-related limitations. 
However, mastering the specificity of liquid nano-NMR is still in its infancy and gaining a fundamental microscopic understanding of quantum sensing at aqueous interfaces is of great importance for the development of the field.

In this paper, we address the mechanism of quantum sensing of the structure of water and simple solvated ions at the interface with a 2D material, using a NV-like spin defect hosted in graphene. Although the electronic properties of graphene are not amenable to host spin qubits, due to the absence of an electronic gap, we consider this material as a model 2D host~\cite{Ye:2019}, for computational simplicity. Using molecular dynamics (MD) simulations and coherence properties calculations, we investigate the effect of water on the spin defect's coherence time, and the sensitivity of the color center's decoherence to the structural and dynamical properties of the liquid at the interface.

The rest of the paper is organized as follows: we first present the methodology adopted in our study (section II), followed by our results (section III) and conclusions (section IV).

\section{Methods}
We performed classical MD simulations of pure water and simple ions dissolved in water in contact with two parallel graphene sheets at a distance of 28 \AA{},
using the non-polarizable TIP4P/2005~\cite{Abascal:2005} water model for the pure liquid  and the SPC/E model~\cite{Romer:2012} for ions in water. (Hereafter we refer to TIP4P/2005 as simply TIP4P). We used periodically repeated supercells with up to 15,000 molecules, carrying out detailed finite size scaling tests, reported in the SI. The water-carbon interactions
were described by the 12-6 Lennard-Jones (LJ) potential  given in Ref.~\cite{Werder:2003}; in the case 
of ions we used the LJ interaction parameters from Refs.~\cite{Williams:2017, Dockal:2019} (see SI for details). All simulations were
carried out in the NVT ensemble with the LAMMPS code~\cite{Thompson:2022} (see SI). 

\begin{figure}[!ht]
    \begin{subfigure}[b]{0.3\linewidth}\label{fig:Cartoon}
        \centering
        \begin{tikzpicture}
        \node[anchor=north west, inner sep=0] (image) at (0, 0)
        {\includegraphics[trim={0.0em 0.0em 0.0em 0.0em},clip,height=4.2cm, width=1.2\textwidth, cfbox=black 1pt 0.2pt]{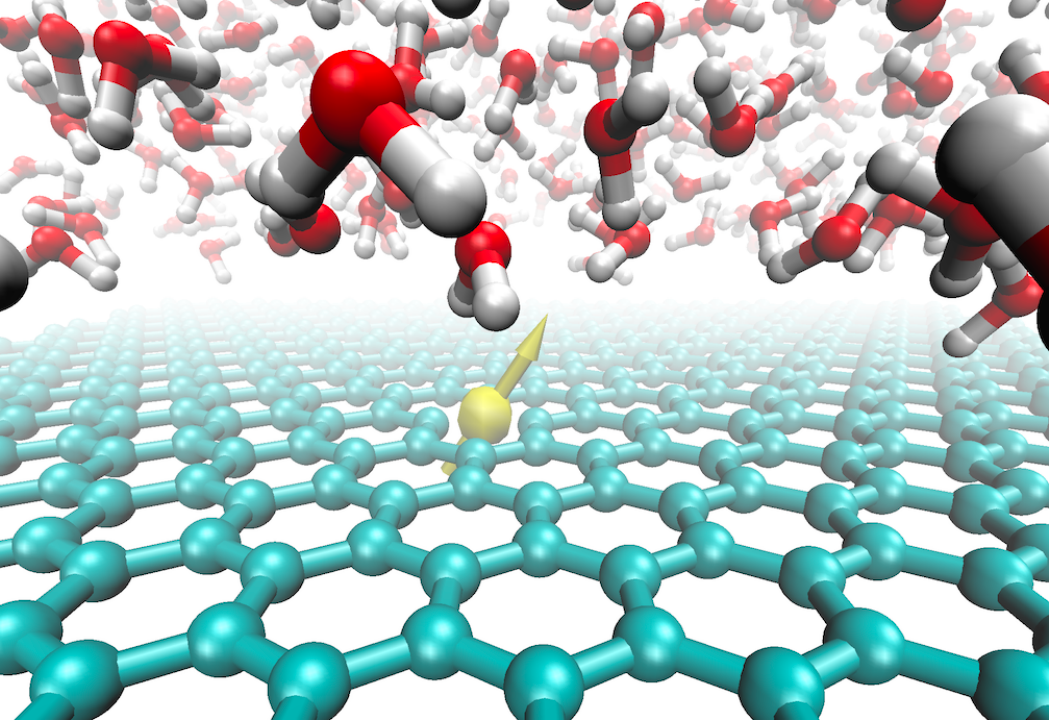}};
        \end{tikzpicture}
    \end{subfigure}     
    \caption{Pictorial representation of a NV-like center (yellow sphere) hosted in a graphene layer; hydrogen, oxygen and carbon atoms are represented by  white, red and cyan spheres, respectively.}
    \label{fig:SYSTEM}
\end{figure}

The coherence function of the NV-like qubit interacting with water, schematically depicted in Fig.~\ref{fig:SYSTEM},  was computed using the cluster correlation expansion method~\cite{Yang:2008, Yang:2009} as implemented in the PyCCE code~\cite{OnizhukPyCCE:2021} (see SI for details). The interaction between the electronic spin and nuclear spin was described using both a semiclassical approximation and a quantum model, which we present next. 

\subsection{\label{sec:SA} Semiclassical Approximation}

We first describe a semiclassical approach to investigate how  decoherence time of a near-surface spin qubit is affected by the protons of water. The nuclear spins are coupled to each other via magnetic dipolar coupling, and the nuclear spin bath interacts with the qubit with spin $S_z$ through hyperfine interactions. 


We use an effective semiclassical Hamiltonian to represent the central spin:


\begin{equation} \label{eq1}
\tilde{H}_{cl}(t) = y(t) B(t) S_{z}
\end{equation}

where the effective magnetic field $B(t)$ is a classical stochastic variable whose action approximates that of the quantum operator $\hat{B_{z}}$. We assume the system to be interrogated by microwave radiation, as customary for spin defects in diamond, and subject to $\pi$-pulse control around the $z$ axis. In a toggling frame, each $\pi$ pulse causes the effective field felt by the central spin to flip, and $y(t)=\pm 1$ changes its sign whenever a $\pi$ pulse is applied. The magnetic field is given by

\begin{equation} \label{eq2}
    B(t) = \sum_{m = 1}^{M} A^m_{zz} \hat{I_z}
\end{equation}

where the sum runs over all $M$ protons and the $z$$z$-component of the hyperfine interaction A$^m_{zz}$. Considering dipole-dipole interactions, the hyperfine tensor of the $m$th proton is given by

\begin{equation} \label{eq3}
    A^{m} = - \gamma_{S} \gamma_{m} \frac{\hbar ^{2}}{4 \pi \mu_0} \left[ \frac{3 \vec{r_{Sm}} \otimes \vec{r_{Sm}}-|r_{Sm}|^{2} I}{|r_{Sm}|^5} \right] 
\end{equation}

where $\gamma_{m}$ ($\gamma_{S}$) is the gyromagnetic ratio of the $m$ proton (central spin) and $\vec{r_{Sm}}$ is the distance between the central spin and the $m$ proton. $I$ is a 3x3 identity matrix. We set $\hbar$ = 1 so that hyperfine interactions have units of frequency and $\mu_0$ is the vacuum magnetic permeability.

If the fluctuating field is Gaussian~\cite{Jacobs:2010}, an assumption valid in the pure dephasing regime \cite{deSousa:2009, Biercuk:2011, Degen:2017, Cywinski:2008} and if the noise can be assumed as arising from the sum of a  large number of classical fluctuating variables (nuclear spins)~\cite{ClerkNoise:2010}, we can write the coherence function of the central spin as:

\begin{equation} \label{eq4}
    \Lb(t) = e^{-\langle\phi^{2}(t)\rangle}
\end{equation}

where the averaged phase squared $\langle \phi^{2} \rangle$ accumulated by the spin qubit is obtained as:

\begin{equation} \label{eq5}
    \langle\phi^{2}(t)\rangle = \int_{0}^{t}  \,d\tau C(\tau)F(\tau) 
\end{equation}

Here F($\tau$) is a filter function \cite{Young:2012} describing the action of the control pulses. Assuming that the main source of magnetic noise in our system is due to protons, C($\tau$) is the time autocorrelation function given by the sum of correlations over all $m$ protons:

\begin{equation} \label{eq6}
    C(\tau) = \frac{1}{4} \sum_{m = 1}^{M} \langle A_{zz}^{m}(\tau)\cdot A_{zz}^{m}(0) \rangle
\end{equation}

These correlations were computed on trajectories generated by  the classical MD simulations described above.  Convergence was achieved with supercells containing at least 10,000 molecules of water (see SI for details), similar to what reported for radial dipole-dipole space correlation functions in Ref. \cite{Zhang:2014}. The decay of C($\tau$) can be  fitted by a stretched exponential function:

\begin{equation} \label{eq7}
    C(\tau) = C(0) + \Delta^2 (e^{-(\tau/\tau_c)^{n_c}}-1)
\end{equation}

where $\Delta$ is the correlation amplitude, $\tau_c$ is the correlation time, and $n_c$ is the stretched exponent.  The phase accumulated by the qubit ($ \langle\phi^{2}(t)\rangle $) and its coherence function ($ \Lb(t)$) were computed with the PyCCE code~\cite{OnizhukPyCCE:2021}. The coherence time T$_2$ is obtained by fitting $\Lb(t)$ to a stretched exponential function: $\Lb(t) = e^{-(\frac{t}{T_2})^n}$.

\subsection{\label{sec:QM} Quantum Model}
At low temperature, where diffusion of water molecules is such that motional narrowing becomes unremarkable, we compared  the results of our semiclassical model with those of a quantum model, where we used 
the following spin Hamiltonian \cite{Seo:2016} to describe an NV center:

\begin{equation} \label{eq9}
\hat{H} = \hat{H}_{S} + \hat{H}_{SB} + \hat{H}_{B}
\end{equation}

where:

\begin{equation} \label{eq10}
\hat{H}_{S} = \mathbf{SDS} + \mathbf{B}\gamma_{S} \mathbf{S}
\end{equation}

\begin{equation} \label{eq11}
\hat{H}_{SB} = \sum_i \mathbf{S} \mathbf{A}_{i} \mathbf{I}_{i}
\end{equation}

and 

\begin{equation} \label{eq12}
\hat{H}_{B} = \sum_i \mathbf{I}_{i} \mathbf{P}_{i} \mathbf{I}_{i} + \mathbf{B}\gamma_{i} \mathbf{I}_{i} + \sum_{i>j} \mathbf{I}_{i} \mathbf{J}_{ij} \mathbf{I}_{j}
\end{equation}

Here $\mathbf{S}$ = ($\hat{S}_x$, $\hat{S}_y$, $\hat{S}_z$) are the components of the spin operators of the central spin, and  $\mathbf{I}$ = ($\hat{I}_x$, $\hat{I}_y$, $\hat{I}_z$) are the components of the bath spin operators; $\mathbf{D(P)}$ is the self-interaction tensor of the central (bath) spin, which corresponds to the zero field splitting (ZFS) tensor for the electronic spin and the quadrupole interactions tensor for the nuclear spins; $\gamma_{i}$ describes the interaction between the $i$-th spin and the external magnetic field $\mathbf{B}$; $\mathbf{A}$ is the hyperfine coupling tensor, describing the interaction  between the central and bath spins; $\mathbf{J}$ is the interaction tensor between bath spins.

To compute the homogeneous dephasing time of the NV center, we first considered the Han echo \cite{Hahn:1950}  pulse sequence, which is  insensitive to energy detunings to first order\cite{Young:2012}. Additional pulse sequences are discussed later in the Results section. The decoherence function of the qubit is then obtained by computing the off-diagonal elements of the reduced density matrix ($\hat{\rho}_{S}$) of the qubit after tracing out the bath degrees of freedom as:

\begin{equation} \label{eq13}
\Lb(t) = \frac{\langle 1| \hat{\rho}_{S}(t) |0\rangle}{\langle 1| \hat{\rho}_{S}(0) |0\rangle}  
\end{equation}

where the coherence function $\Lb(t)$ characterizes the loss of the relative phase of the $|0\rangle$ and $|1\rangle$ levels (see SI for details).

We found that second order correlations were sufficient to converge the calculation of the coherence function, indicating that the  main contribution to decoherence comes from pairwise nuclear transitions induced by
nuclear dipole-dipole couplings.  In our calculations the size of the bath was truncated at 15 \AA, and the dipole cutoff at 6 \AA, as usually done when studying NV centers in diamond (tests as a function of these cutoffs are reported in the SI). With these parameters we determined that cells with 1024 molecules were sufficiently large to carry out calculations with the quantum model. The value of T$_2$ is then obtained by fitting $\Lb(t)$ as indicated in the case of semiclassical model ($\Lb(t) = e^{-(\frac{t}{T_2})^n}$).  


\section {Results}

We now turn to describing our results as a function of several physical parameters characterizing the water/graphene interface, starting from the temperature of the liquid, followed by the strength of the surface-water interaction and the presence of ions.
\subsubsection{\label{sec:SA_Temperature} Temperature Effects}
\begin{figure}[!ht]
    \begin{subfigure}[b]{0.3\linewidth}\label{fig:T2_Dist}
        \centering
        \begin{tikzpicture}
        \node[anchor=north west, inner sep=0] (image) at (0, 0)
        {\includegraphics[trim={0.0em 0.0em 0.0em 0.0em},clip,height=4.2cm, width=1.0\textwidth]{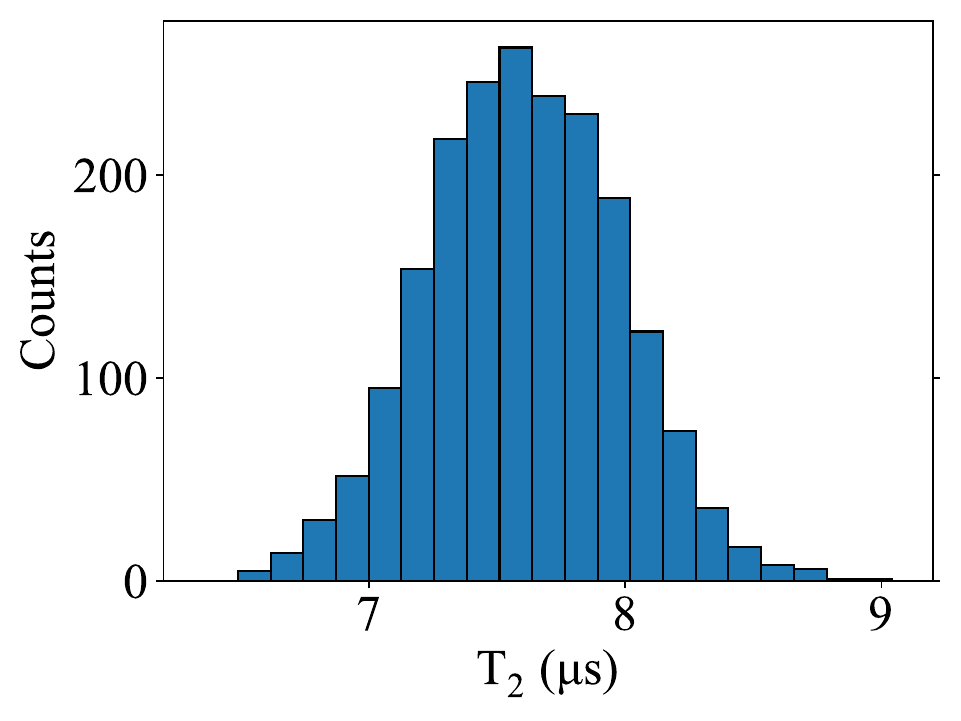}};
        \end{tikzpicture}
    \end{subfigure}     
    \caption{Distribution of coherence times computed with the quantum model for the NV-like center in the presence of a-H$_2$O. The computed  average is T$_2$ = 7.6 $\pm$ 0.38 $,\mu$s obtained over 2000 configurations extracted from a single molecular dynamics simulation trajectory.}
    \label{fig:T2_DIST}
\end{figure}

In Table \ref{table:T2_COMP}, we show results obtained with the semiclassical approximation for the coherence time of the NV center at two temperatures, corresponding to ambient conditions and to a supercooled regime (250K) close to the freezing point of water (with the TIP4P potential adopted here).  We also include results derived with the quantum model for amorphous ice, with an histogram shown in Fig.~\ref{fig:T2_DIST}, whose results have been obtained by carrying calculations for 2000 snapshots extracted from MD simulations of liquid water at T = 300 K and by averaging the data.  The quantum model does not account for the dynamics of the protons, which in each configuration are frozen. For this reason we called the structural model used for the quantum calculations amorphous ice (a-H$_2$O)(see SI for details). We find that due to motional narrowing\cite{Slichter:1990}, in the liquid the decoherence time of the qubit increases by more than an order of magnitude from the supercooled regime to ambient conditions. As the self-diffusion coefficient of water decreases with T, in the supercooled regime, T$_2$ reaches a value similar to that of amorphous ice, showing good agreement between the quantum model and the semiclassical approximation.
Note that, using the quantum model,  T$_2$ computed for a qubit in monolayer graphite in the absence of water (T$_2$ = 8.72 $\pm$ 1.32 $\mu$s) is in agreement with the value reported by Ref. \cite{Ye:2019} (T$_2$ = 9.57 $\mu$s); in addition, our results for  a-H$_2$O are consistent with those for an  electron spin in a single wall carbon nanotube (SWCNT) \cite{Chen:2023}, where  it was shown that the presence of a $^1$H nuclear spin bath from toluene significantly reduced the coherence time of a defect in the SWCNT to 8.7 $\mu$s . 

The motional narrowing effect observed or liquid water is a well known phenomenon in NMR: the motion of water molecules causes faster fluctuations of the magnetic noise than in ice, leading to longer dephasing time or, in terms of the Fourier transform of $ \Lb(t)$, to a narrower magnetic resonance line of the central spin. In other words, the qubit does not acquire  a sufficiently large  phase to decohere due to the fast fluctuating magnetic noise of the environment. According to Eq.~\ref{eq7}, the characteristic timescale within which a decay of correlations is observed is $\tau_c$; for water $\tau_c$is on the order of picoseconds, as discussed below,  confirming the rapidly fluctuating magnetic noise, which limits the interaction   of the qubit with the protons of water. 
Interestingly, the motional narrowing principle has been recently used to protect and even extend the coherence of near-surface NV centers by coherent radio-frequency driving of surface electronic spins \cite{Bluvstein:2019} and polychromatic driving of the surface-spin bath \cite{Joss:2022}. 

\begin{figure}[!ht]
    \begin{subfigure}[b]{0.3\linewidth}\label{fig:SA_TEMP_L}
        \centering
        \begin{tikzpicture}
        \node[anchor=south west, inner sep=0] (image) at (0,0)
        {\includegraphics[trim={0.0em 0.0em 0.0em 0.0em},clip,height=4.2cm, width=1.0\textwidth]{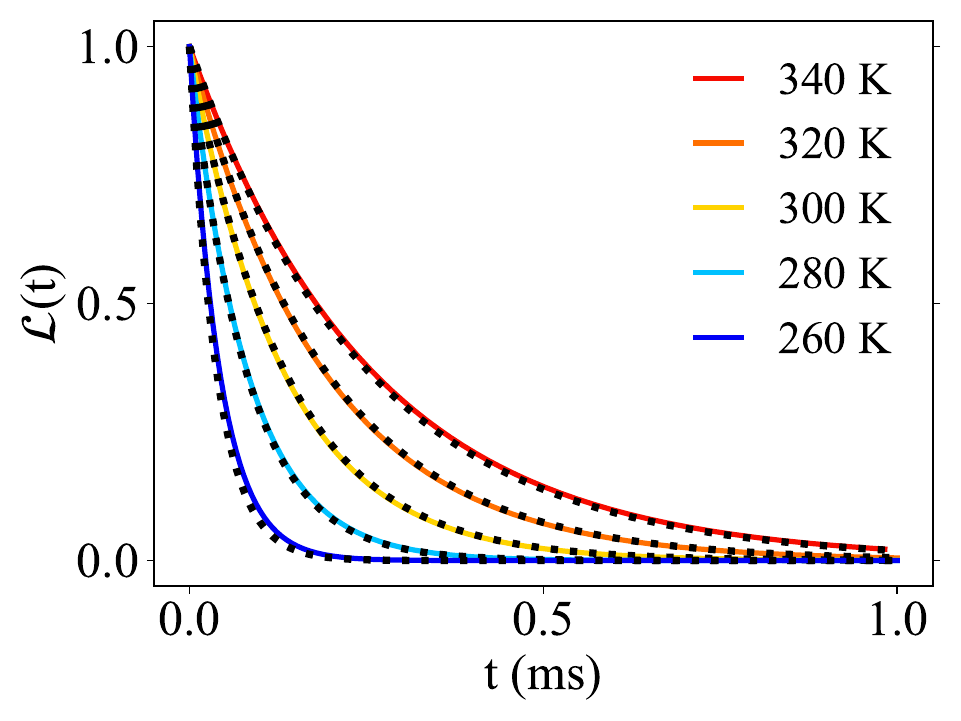}};
        \node[anchor=north west] at ([xshift=-9pt,yshift=5pt]image.north west) {(a)};
        \end{tikzpicture}
    \end{subfigure}    
    \hspace{0.5em}
    \begin{subfigure}[b]{0.3\linewidth}\label{fig:T2_VS_TC}
        \centering
        \begin{tikzpicture}
        \node[anchor=south west, inner sep=0] (image) at (0,0)
        {\includegraphics[trim={0.0em 0.0em 0.0em 0.0em},clip,height=4.2cm,width=1.07\textwidth]{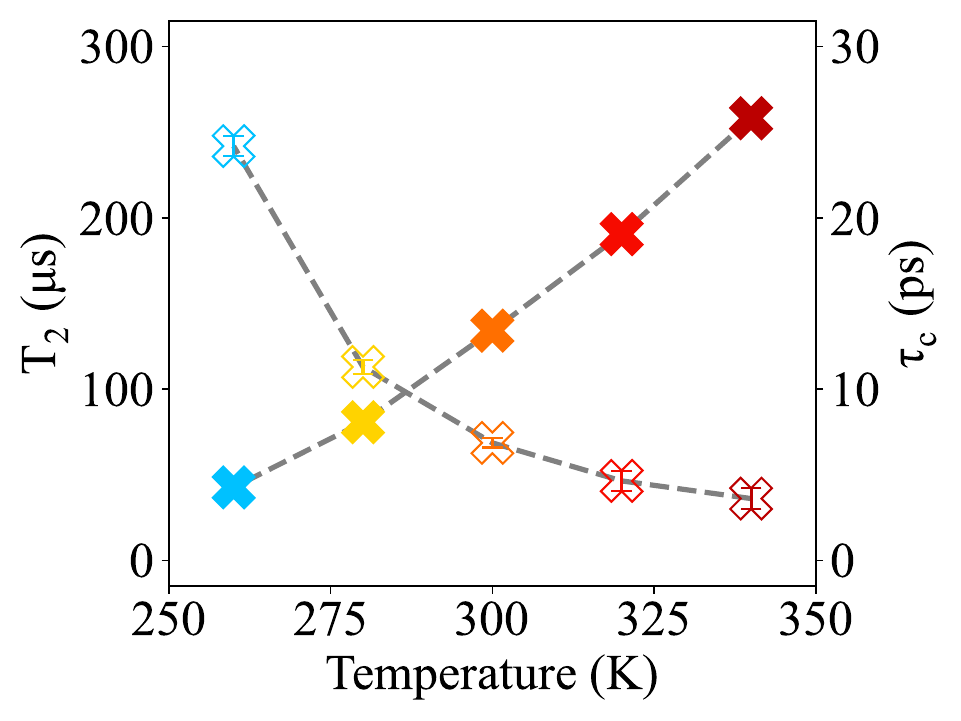}};
        \node[anchor=north west] at ([xshift=-9pt,yshift=5pt]image.north west) {(b)};
        \end{tikzpicture}
    \end{subfigure}  
    \hspace{0.5em}
    \begin{subfigure}[b]{0.3\linewidth}\label{fig:DXY_VS_TEMP}
        \centering 
        \begin{tikzpicture}
        \node[anchor=south west, inner sep=0] (image) at (0,0)
        {\includegraphics[trim={0.0em 0.0em 0.0em 0.0em},clip,height=4.2cm, width=1.0\textwidth]{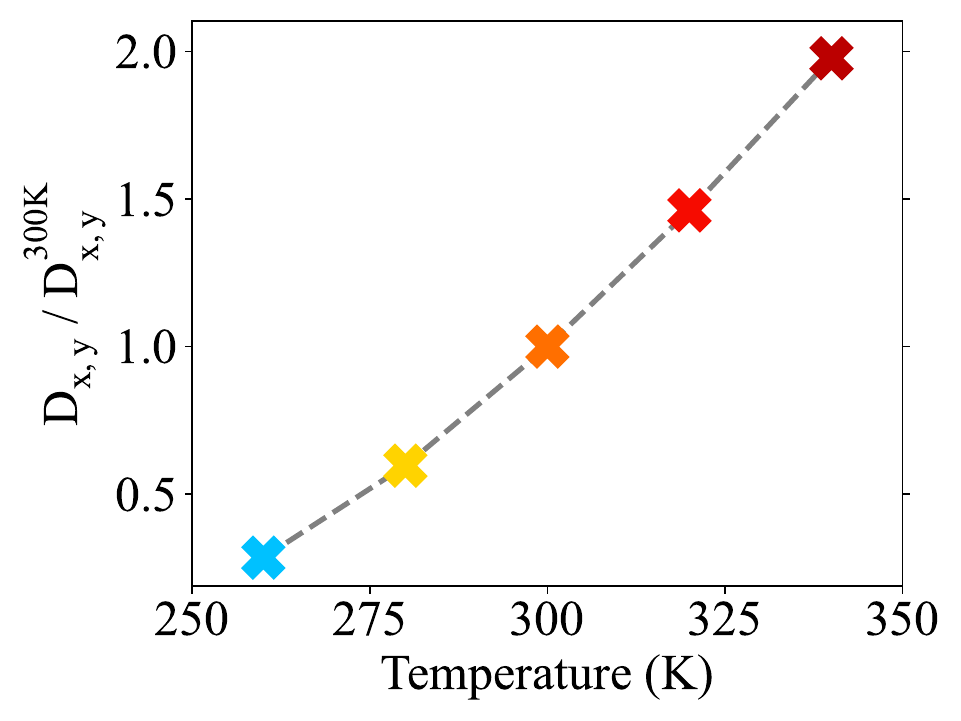}};
        \node[anchor=north west] at ([xshift=-9pt,yshift=5pt]image.north west) {(c)};
        \end{tikzpicture}
    \end{subfigure}   
    \caption{(a) The coherence function $\Lb(t)$ of the NV-like center obtained by computing the spin's phase $\langle \phi^{2} \rangle$ (Eq.~\ref{eq4}) and by  using Eq.~\ref{eq14}, shown as solid and dotted lines, respectively. (b) Coherence times T$_2$ (filled symbols) and correlation times $\tau_c$ (empty symbols), and (c) planar diffusion coefficient D$_{x,y}$ ratio in the plane parallel to the surface, relative to that of water at 300 K as a function of temperature. Dashed solid lines are used to guide the eye.}
    \label{fig:TEMPS}
\end{figure}
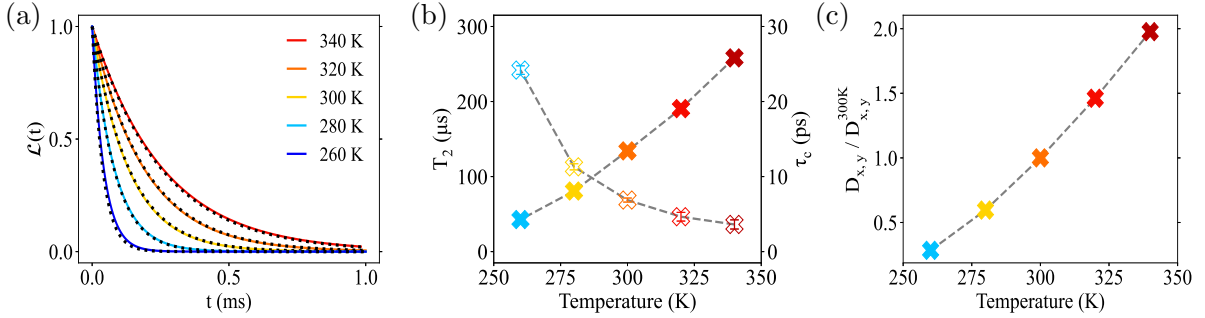

\begin{table}
    \begin{center}
    \caption{Comparison of coherence times (T$_2$) computed with two different approaches (see text) for water and amorphous ice. For liquid water we report simulations carried out with 15,000 molecules and values averaged over 3 independent (20 ns) MD runs. For ice we report the average over 2000 configurations extracted from a single (2 ns) molecular dynamics simulation (see text).}
    \begin{tabular}{cccc}
    \hline
    \hline
    System & Approach & Temperature, K & T$_2$, $\mu$s \\ [3pt]
    \hline
    Amorphous Ice & Quantum Model & $\sim$0 & 7.6 $\pm$ 0.38 \\
    Liquid Water & Semiclassical Approx. & 300 & 117.33 $\pm$ 2.02 \\
    Supercooled Water & Semiclassical Approx. & $\sim$250 & 8.08 $\pm$ 3.67 \\
    \hline
    \hline
    \end{tabular}
    \label{table:T2_COMP}
    \end{center}
\end{table}

To illustrate in detail the impact of the temperature on the motional narrowing effect, we performed MD simulations with 15,000 water molecules at different temperatures and then computed T$_2$. In Fig.~\ref{fig:TEMPS} (a) we show the decay of the coherence function (Eqs.~\ref{eq4} and ~\ref{eq14}) as a function of time (t) and in Fig.~\ref{fig:TEMPS} (b) T$_2$ and $\tau_c$  as a function of temperature. To the best of our knowledge there is no available data on NMR of water protons obtained via NV centers, and hence a direct comparison with experiments is unfortunately not yet possible. 

\begin{figure}[!ht]
    \begin{subfigure}[b]{0.3\linewidth}\label{fig:SLOW_FAST_NOISE}
        \centering
        \begin{tikzpicture}
        \node[anchor=south west, inner sep=0] (image) at (0,0)
        {\includegraphics[trim={0.0em 0.0em 0.0em 0.0em},clip,height=4.2cm,width=1.0\textwidth]{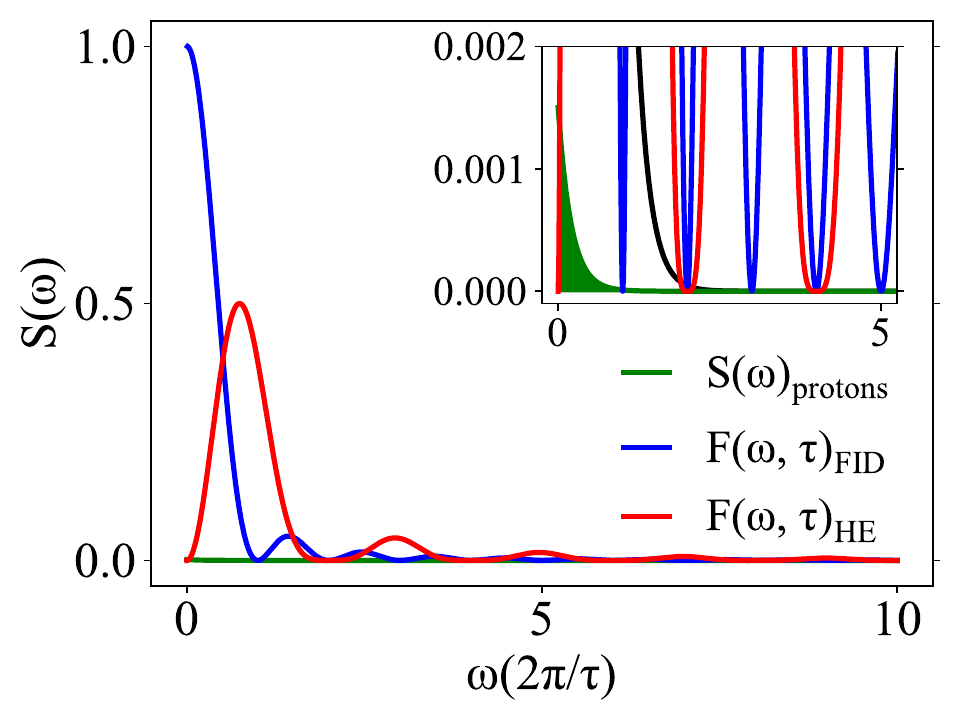}};
        \end{tikzpicture}
    \end{subfigure}
    \caption{Power spectral density ($S(\omega)_{protons}$) of the fast fluctuating magnetic noise induced by protons in water, and Ramsey ($F(\omega)_{FID}$) and Hahn echo (($F(\omega)_{HE}$)) filter functions. Shaded areas in the inset show the phase accumulated by the qubit under an Hahn echo pulse sequence.}
    \label{fig:NOISE}
\end{figure}

We note that the value of $\tau_c$ is always on the order of picoseconds, similar to characteristic timescales of bulk water such as dipole moment/OH bond reorientation times and average lifetime of hydrogen bonds. 
Since  $\tau_c$ $\ll$ T$_2$, we can approximate the qubit's exponentially decaying coherence function at t $\gg$ $\tau_c$ with~\cite{Yang_Review:2017}:

\begin{equation} \label{eq14}
    T_2 \approx \frac{1}{b_{rms}^2 \tau_c}
\end{equation}

by noting that only values $\lvert{t_1-t_2}\rvert$ $\lesssim$ $\tau_c$ contribute to the integral in Eq.~\ref{eq5}. Here, ${b_{rms}^2}$ = C(t=0). As shown in Fig.~\ref{fig:TEMPS} (a) by dotted lines, our computed results agree quite well with the predictions of Eq.~\ref{eq14}. This means that dynamical decoupling sequences should have no effect on the measurement of T$_2$. We confirmed this finding by performing calculations of the phase accumulated by the qubit with a Ramsey sequence (DD$_p$=0), Hanh echo (DD$_p$=1), and DD$_p$=2, 3,  4,  obtaining identical results. To illustrate this finding, in Fig.~\ref{fig:NOISE} we report the power spectral density S($\omega$)$_{protons}$ given by the Fourier transform of the autocorrelation function~\cite{ClerkNoise:2010} in Eq.~\ref{eq6}:

\begin{equation} \label{eq15}
    S(\omega)_{protons} = \int e^{i\omega\tau}C(\tau)d\tau
\end{equation}

where the filter functions entering the expression of the $C$ function, and corresponding to Ramsey (FID) and Hahn echo (HE) pulse sequences are respectively defined as~\cite{Yang_Review:2017}:

\begin{equation} \label{eq16}
    F(\omega, \tau)_{FID} = \frac{sin^2(\omega \tau/2)}{(\omega \tau/2)^2}
\end{equation}

\begin{equation} \label{eq17}
    F(\omega, \tau)_{HE} = \frac{sin^4(\omega \tau/4)}{(\omega \tau/4)^2}
\end{equation}

From the expression of the filter functions, one expects the low-frequency noise to be removed by the dynamical decoupling sequence. However we show in the inset of Fig.~\ref{fig:NOISE} that in the case of diffusing protons,  the low-frequency noise cannot be removed from the fast fluctuating magnetic noise, i.e. the phase accumulated by the qubit and filtered by two different switching functions (FID or Hahn echo) is essentially the same due to the low and broad noise spectrum S($\omega$)$_{protons}$. This result explains the insensitivity of the qubit's coherence time to different filter functions.

\subsection{\label{sec:SA_Hydrophobicity} Strength of the water- surface interaction}
\begin{figure}[!ht]
    \begin{subfigure}[b]{0.3\linewidth}\label{fig:EPSILONS_DENS}
        \centering
        \begin{tikzpicture}
        \node[anchor=south west, inner sep=0] (image) at (0,0)
        {\includegraphics[trim={0.0em 0.0em 0.0em 0.0em},clip,height = 4.2cm,width=1.0\textwidth]{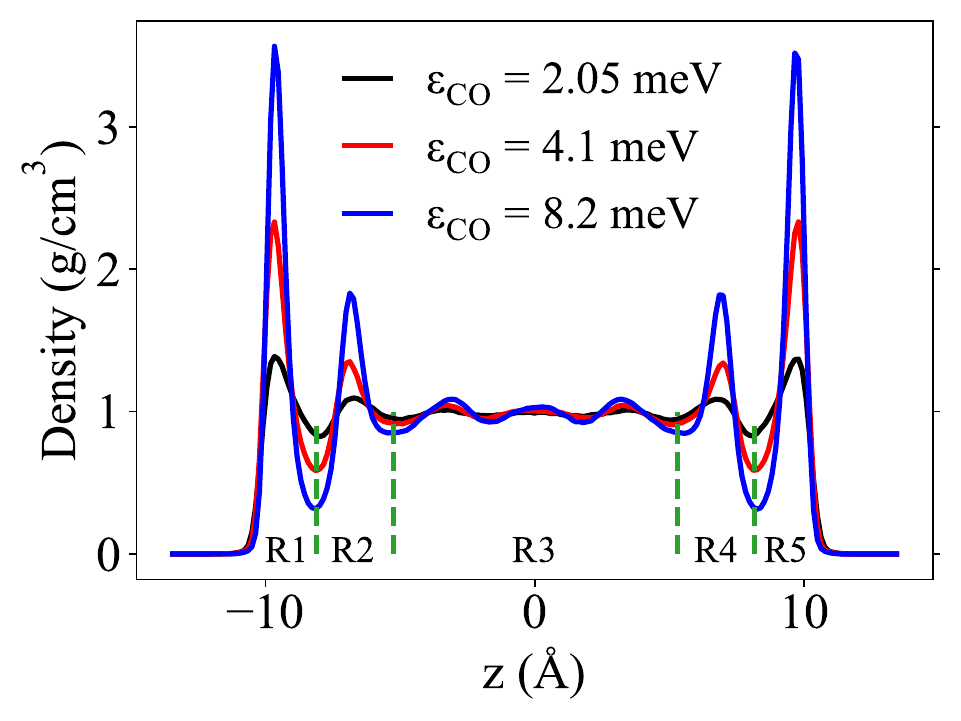}};
        \node[anchor=north west] at ([xshift=-9pt,yshift=5pt]image.north west) {(a)};
        \end{tikzpicture}
    \end{subfigure}    
    \hspace{3em}
    \begin{subfigure}[b]{0.3\linewidth}\label{fig:EPSILONS_L}
        \centering
        \begin{tikzpicture}
        \node[anchor=south west, inner sep=0] (image) at (0,0)
        {\includegraphics[trim={0.0em 0.0em 0.0em 0.0em},clip,height = 4.2cm,width=1.0\textwidth]{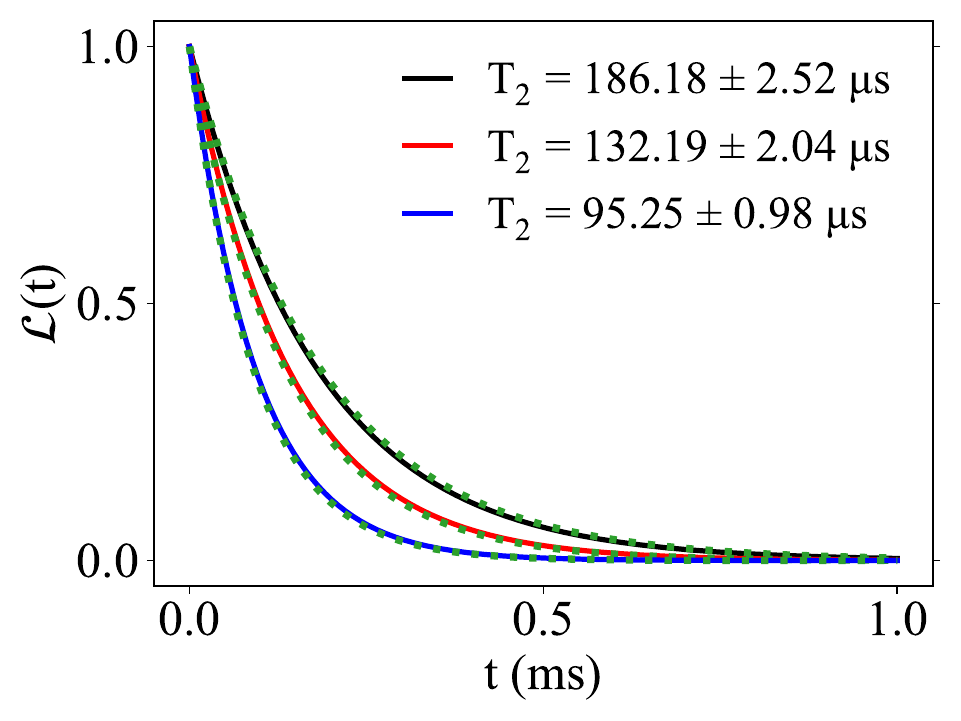}};
        \node[anchor=north west] at ([xshift=-9pt,yshift=5pt]image.north west) {(b)};
        \end{tikzpicture}
    \end{subfigure} 
    \caption{(a) Density distribution of water at T= 300 K, projected along the $z$-axis, computed for different strengths of the surface-water interaction ($\epsilon_{CO}$; see text). We defined three different regions in the liquid, which is placed between two symmetric graphene layers: an interfacial region (R1 and R5) corresponding to distances over which the density of the liquid is sensibly different from that of bulk water; a central region (R3) where the density of the liquid is the same as in the bulk; and an intermediate region (R2 and R4) with density variations  much smaller than in R1 and R5 but still larger than in R3; these regions  are indicated by dashed green lines. (b) Coherence function $\Lb(t)$ of the qubit (Eq.~\ref{eq4}) computed for the three different LJ parameters $\epsilon_{CO}$ given in (a). We also show  the decoherence function computed with Eq.~\ref{eq14} by  dotted lines.}
    \label{fig:EPSILONS}
\end{figure}

Another interesting variable to investigate is the  effect of water-surface interactions on coherence times. To do so, we carried out MD simulations with different values of the LJ parameters chosen for the water-graphene interactions, using both the semiclassical and  quantum model.  A change in the strength of the interaction not only affects the density of the first layer of water (Fig.~\ref{fig:EPSILONS} (a)) but also modifies dipolar fluctuations along the $z$ direction perpendicular to the surface\cite{Zhang:2013}. In Fig.~\ref{fig:EPSILONS} (b) we show that the qubit's coherence function decays faster when the density of the first layer of water is increased. An increase in density leads to  a longer qubit-proton interaction,  that in turn reduces motional narrowing and hence reduces the coherence time. In the case of a-H$_2$O, where motional narrowing is absent, the dependnece of T$_2$ on the liquid density is found to be negligible (see SI).

We characterized the reorientation dynamics at the interface by computing the water dipole ($\tau_{DM}$) and OH bond ($\tau_{OH}$) reorientation times, the average lifetime ($\tau_{HB}$) and average number ($\langle n_{HB}\rangle$) of hydrogen bonds in specific regions of the liquid (R1-R5), as defined  in Fig. \ref{fig:EPSILONS} (a) by dashed vertical lines (see details in Tables SIII-SIV). We note that, although the strength of the water-graphene interaction (as represented by the LJ parameter $\epsilon_{CO}$) does not substantially affect the values of  $\langle n_{HB}\rangle$ and $\tau_{HB}$ in the interfacial region, relative to the central, bulk region (R3), the values of $\tau_{DM}$ and $\tau_{OH}$ do vary in the interfacial regions R1 and R5. These values decrease by approximately $\sim$10\% with $\epsilon_{CO}$=2.05 meV, and  increase by $\sim$20\% with $\epsilon_{CO}$=8.2 meV when compared to the  result obtained  with the reference value $\epsilon_{CO}$=4.1 meV~\cite{Werder:2003} used for all simulations reported in Fig.3. 


Hence, increasing the strength of the surface-water interaction not only enhances the alignment of  water molecules parallel to the surface \cite{Zhang:2013} but also reduces the reorientation time  at the graphene-water interface, which in turn reduces the motional narrowing effect, thus allowing the qubit to increase its phase accumulation and to decohere over shorter timescales: the T$_2$ value is reduced by $\sim$30\% with the highest value of $\epsilon_{CO}$ (8.2 meV). In summary, the stronger is the surface-water interaction, the shorter is T$_2$. This result holds in the presence of motional narrowing, while in the case of a-H$_2$O, we find that there is hardly any effect of the strength of this interaction on T$_2$ (see SI for details).

We also performed additional calculations of T$_2$ for a confinement distance  h$_{conf}$ $\simeq$ 15 \AA{}, i.e. approximately a factor of 2 smaller than that considered for all simulations discussed above.  We found that computed T$_2$ values depend weakly on this variable, showing again that the major role in determining the decoherence of the qubit is played by  properties of water in the interfacial region.

\subsection{\label{sec:SA_Ions} Presence of ions}
 We now turn to elucidate the role of ions in  determining the coherence time of the qubit.
 
 
 Here, we use the LJ interaction parameters from Refs. \cite{Williams:2017, Dockal:2019} to account for the surface polarization of graphene sheets  in the presence of ions in the system, and the SPC/E force field (FF). We adopt the  SPC/E and not the TIP4P FF in this part of the work because it is only for the former one that interaction parameters of alkali halides and alkaline-earth halides with water and graphene have been derived in the literature.
Hence, before discussing our results on coherence times for ions, we carried out a comparison of  T$_2$ computed for pure water with  TIP4P and SPC/E. The experimental diffusion coefficient at room T is overestimated by both FF's, but to a different extent: $\sim$4\% and $\sim$20\% by TIP4P and SPC/E, respectively. As expected, the value of T$_2$ computed using SPC/E MD trajectories is larger (by $\sim$30\%) than that obtained with TIP4P/2005.  This is a clear consequence of the difference between the diffusion coefficients computed with the two FFs and to the enhancement of the motional narrowing effect in the case of SPC/E, extending the lifetime of the qubit. To have a fair comparison between pure water and salty water, from here on, we will compare all the results obtained for  ions to the results of water obtained with the SPC/E force field.

\begin{figure}[!ht]
    \begin{subfigure}[b]{0.3\linewidth}\label{fig:T2_ALL_IONS}
        \centering
        \begin{tikzpicture}
        \node[anchor=south west, inner sep=0] (image) at (0,0)
        {\includegraphics[trim={0.0em 0.0em 0.0em 0.0em},clip,height = 4.2cm,width=1.0\textwidth]{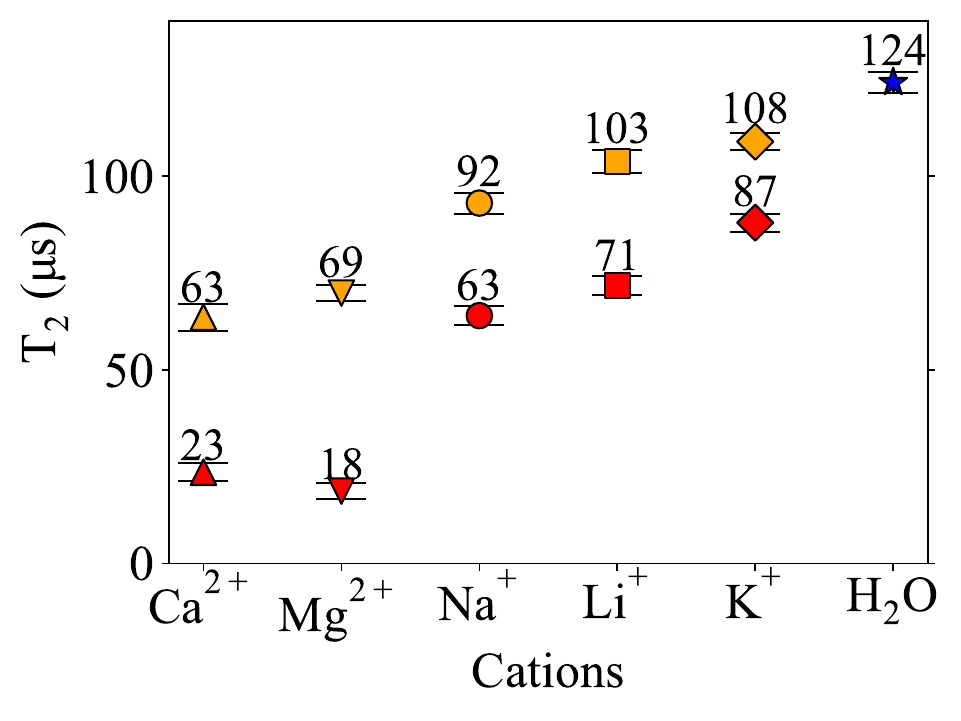}};
        \node[anchor=north west] at ([xshift=-9pt,yshift=5pt]image.north west) {(a)};
        \end{tikzpicture}
    \end{subfigure}
    \hspace{3em}
    \begin{subfigure}[b]{0.3\linewidth}\label{fig:T2_VS_TAU_C}
        \centering
        \begin{tikzpicture}
        \node[anchor=south west, inner sep=0] (image) at (0,0)
        {\includegraphics[trim={0.0em 0.0em 0.0em 0.0em},clip,height = 4.2cm,width=1.0\textwidth]{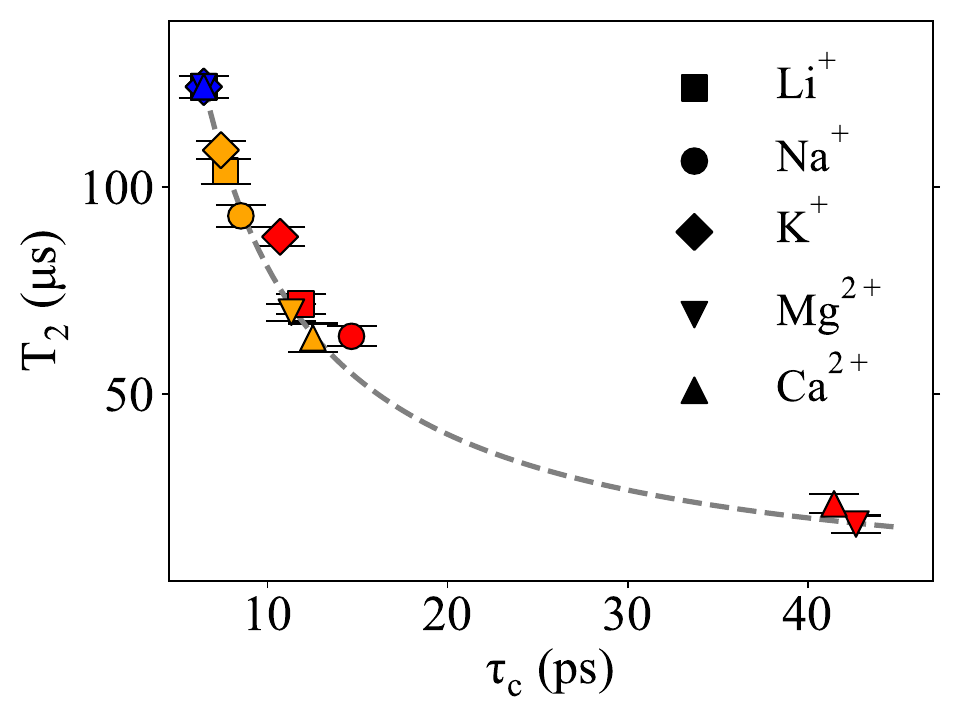}};
        \node[anchor=north west] at ([xshift=-9pt,yshift=5pt]image.north west) {(b)};
        \end{tikzpicture}
    \end{subfigure} 
    \caption{(a) Coherence times T$_2$ for all ions investigated in this work and (b) T$_2$ as a function of correlation time $\tau_c$. Results for pure water, low (0.5 M) and high (2.5 and 1.5 M for mono-valent and divalent ions, respectively) concentrations of salt are represented by blue, orange, and red filled markers, respectively. MD simulations were conducted at 300 K. Dashed solid line is used to guide the eye.}
    \label{fig:IONS}
\end{figure}

Fig.~\ref{fig:IONS} (a) shows the computed values of T$_2$ for all cations considered here (monovalent Li, Na and K and divalent Mg and Ca). In all cases we considered solutions with the same anion (Cl) and we investigated two concentrations 0.5 M (orange filled markers) and 2.5 and 1.5 M (red filled markers) for monovalent and divalent species, respectively. As shown in the SI (Fig. S9 (a)),  all cations considered in this work have a strong preference to reside close to the surface. Fig.~\ref{fig:IONS} (b) shows the T$_2$ values as a function of the correlation time $\tau_c$ (see Eq.~\ref{eq7}). We find that T$_2$ depends on the dynamical properties of the ion solvation shells: the longest the residence time of water molecules in the first solvation shell, the more the motional narrowing is affected and the shortest the coherence time. To understand these dependencies we computed the average lifetime of hydrogen bonds and water dipole moment in the interfacial regions (see SI) and found that they increase relative to pure water and they increase in going from Li, to Na and K and further increase for Ca and are the largest for Mg, consistently with what known about the 'stiffness' of the solvation shells for these ions. This is an important result as it shows that near-surface spin qubits can in principle discriminate between different types of ions present at the interface and their coherence time depends on the concentration of the ions. Interestingly at all concentration the computed T$_2$ follows an inverse dependence on $\tau_c$ (Fig.~\ref{fig:IONS}). We find that the exponents $n_c$ of Eq.~\ref{eq7} entering the expression of the correlation function $C(\tau)$ show a weak variation as a function of the cation and even the concentration, similar to the parameter $\Delta$ as shown in Table SXIII). Instead $\tau_c$ shows remarkable variations in going from monovalent to divalent ions and as a function of concentration.




\subsection{\label{sec:Conclusions} Conclusions}
In summary, we investigated the interaction of an NV-like spin qubit with a model interface between a 2D material and water and simple aqueous solutions. We showed that the decoherence time of the qubit is sensitive to the microscopic properties of the liquid at the interface, including temperature, the strength of the surface-water interaction and the dynamical properties of the solvation shells of the ions residing close to the solid-liquid interface. We find that the sensitivity of the qubit (on the order of microseconds),  stemming from motional narrowing effects, is inversely proportional to characteristic 
time scales of hybrogen bonding and dipole correlations in water, which are of the order of ps. We also find that motional narrowing effects are responsible for the insensitivity of the decoherence time to dynamical decoupling sequences in coherence measurements.

Our results point at the exciting perspective of near surface quantum sensors to investigate aqueous interfaces at the microscopic scale, and at the future powerful interplay between simulations and experiments for such studies. In this work we could only address interfaces between water and a 2D material where no chemical reactions occur, due to the use of simple force-fields, chosen because of the large scale simulations required (up to 15,000 water molecules) and long time scales (tens of ns for each of the simulations we conducted). However, one can envision that with the advent of increasingly accurate machine-learned FF, as well as of techniques to identify transition states and overall intermediate configurations of chemical reactions occurring at surfaces, one may be able to use quantum sensors to understand reaction pathways, and to use simulations to guide experiments in their identifications. In addition,  one can envision carrying out simulations for multiple substrates and varied spin defects, as a function of additional ions (both cations and anions) and exploring possible universal relations between decoherence time of the qubit and correlation times ($\tau_c$).

The physical principle demonstrated here could also be applied to realize quantum-enhanced sensing platforms for decoherence-based measurements of near-wall flow to better understand, for example, the nature of momentum transfer, the no-slip boundary condition, and the turbulence transition affected by interfacial liquid properties in nano- and micro-fluidic devices. This opens interesting opportunities for the combination of rheological measurements and nano-NMR in surface science.

We close by noting that many questions remain open, including the contributions to the coherence time of terms beyond dipolar interactions and the interplay between magnetic and electric noise at the surface, which could be included in future simulations by computing zero-field splitting and quadrupole parameters from first principles.

\begin{acknowledgments}
A.C., M.O., and G.G. acknowledge the support of NSF QuBBE Quantum Leap Challenge Institute (Grant No. NSF OMA-2121044). We acknowledge the use of the computational facilities (Research Computing Center) at the University of Chicago. The work of GRP-L and the development of the PyCCE code used in this work was supported by the Computational Materials Science Center Midwest Integrated Center for Computational Materials (MICCoM). MICCoM is part of the Computational Materials Sciences Program funded by the US Department of Energy, Office of Science, Basic Energy Sciences, Materials Sciences, and Engineering Division through the Argonne National Laboratory.
\end{acknowledgments}

\section*{Data Availability Statement}
The data supporting our findings are available within the paper and the Supporting Information. Additional relevant data will be made available through Qresp. The codes used in the simulations are all open source.



\clearpage

\section*{\textbf{Supporting Information for "Probing aqueous interfaces with spin defects"}}%
\section*{\label{sec:Methods} Methods}
\subsection*{Classical Molecular Dynamics Simulations} \label{METHODS_MD}
We performed classical molecular dynamics (MD) simulations of water and ions in water between two parallel graphene sheets using the non-polarizable TIP4P/2005~\cite{Abascal:2005} water model and the SPEC/E water model~\cite{Romer:2012}. Periodic boundary conditions were applied along $x$, $y$, and $z$.  The water-carbon interactions was described  by a 12-6 Lennard-Jones potential with parameters $\sigma_{CO}$ and $\epsilon_{CO}$ as proposed by Werder et al. \cite{Werder:2003}. The effect of the magnitude of these parameters on computed coherence times is  discussed in the main text. In the case of ions we used the LJ interaction parameters from Refs. \cite{Williams:2017, Dockal:2019}  and the SPC/E force field (FF). All simulations were carried out in the NVT ensemble  with a time step of 1 fs. Long-range electrostatic interactions were treated by particle-particle-particle-mesh/TIP4P with a precision of 10$^{-4}$. Each simulation was equilibrated for at least 3 ns in the NPT ensemble  and 3 ns in the NVT ensembles before collecting statistics for 20-40 ns every 1 ps. The initial configurations were prepared with Packmol~\cite{Martinez:2009}, VMD~\cite{Humphrey:1996}, and TopoTools~\cite{Kohlmeyer:2019} to generate the input file for the LAMMPS code \cite{Thompson:2022}.

\subsection*{Cluster Correlation Expansion Calculations} \label{METHODS_PYCCE}
To compute the coherence function of the spin defect, we employed the cluster correlation expansion (CCE) method~\cite{Yang:2008,Yang:2009}.This method is one of the most widely used approaches to simulate the quantum decoherence dynamics of spin qubits in a finite spin bath, and  it has provided accurate results for numerous systems~\cite{Seo:2016,Onizhuk1:2021}. The CCE method has also been used to predict properties of  2D platforms for spin qubits~\cite{Ye:2019}, and to conduct a general screening of potential qubit hosts over a wide range of materials~\cite{Kanai:2021}. In this work, we used the conventional CCE method ~\cite{Yang:2008,Yang:2009} as implemented in the PyCCE code~\cite{OnizhukPyCCE:2021}. The coherence function $\Lb(t)$ is factorized into contributions from independent nuclear spin clusters of different sizes \cite{Yang:2008}:

\begin{equation} \label{eqS6}
\Lb(t) = \prod_{i} \tilde{L}^{i} \prod_{i, j}  \tilde{L}^{i,j}... 
\end{equation}

The contributions of different clusters $C$ are computed recursively as:

\begin{equation} \label{eqS7}
\tilde{L}_{c} = \frac{L_C}{\prod_{C'}\tilde{L}_{C' \subset C}} 
\end{equation}

where the subscript $C'$ indicates all subclusters of $C$, and 

\begin{equation} \label{eqS8}
L_{c} = \langle C| \hat{U}^{(0)}_C (t) \hat{U}^{(1)\dagger}_C (t) | C\rangle
\end{equation}

where $|C\rangle$ is the initial state of the cluster $C$; $\hat{U}^{(\alpha)}_C (t)$ is the time propagator defined in terms of the effective Hamiltonian $\hat{H}^{(\alpha)}_C$ conditioned on the levels of the qubit, which  up to second-order perturbation theory, can be written as 

\begin{equation} \label{eqS9}
\hat{H}^{(\alpha)}_C = \langle \alpha | \hat{H}_C |\alpha\rangle + \sum_{i\neq \alpha}\frac{\langle \alpha | \hat{H}_b |i\rangle \langle i | \hat{H}_b |\alpha\rangle}{E_{\alpha}-E_{i}}
\end{equation}

where $|\alpha\rangle$, $|i\rangle$ are eigenstates of the central spin Hamiltonian $\hat{H}_{env}$, $\hat{H}_{C}$ is the Hamiltonian in Eq. 1 in the main text including only the bath spins in the cluster $C$:

\begin{equation} \label{eqS10}
\hat{H}_{C} = \hat{H}_{env}+\hat{H}_{env-b}^{(i\in C)}+\hat{H}_{b}^{(i, j\in C)}
\end{equation}

The point dipole approximation was used to compute hyperfine interactions. All calculations were performed considering conditions similar to those used for Han echo experiments in diamond \cite{Hahn:1950, Seo:2016} (all $\pi$ pulses in the Hanh echo dynamical decoupling sequence are assumed to be ideal, instantaneous, and selective to the central spin). 

The coherence time T$_2$ was obtained by fitting $\Lb(t)$ to a stretched exponential function exp${^-(\frac{t}{T_2})^n}$.

\subsection*{Convergence of CCE} \label{QM_CONV}
We checked the convergence of CCE calculations with respect to the correlation order, the size of the bath (cutoff radius), and the dipole cutoff radius. In Fig.~\ref{fig:CCE}(a) we find that the computed coherence time is numerically converged at the CCE-2 level of theory since higher-order correlations do not yield significant corrections. The size of the bath was truncated at 15 \AA{}, and the dipole cutoff at 6 \AA{} (these also corresponds to value usually adopted to simulate NV centers in diamond). 

\renewcommand{\thefigure}{S1}
\begin{figure}[!ht]
    \begin{subfigure}[b]{0.3\linewidth}\label{fig:CCE_ORDER}
        \centering
        \begin{tikzpicture}
        \node[anchor=south west, inner sep=0] (image) at (0,0)
        {\includegraphics[trim={0.0em 0.0em 20.4cm 0.0em},clip,height=4.2cm, width=1.0\textwidth]{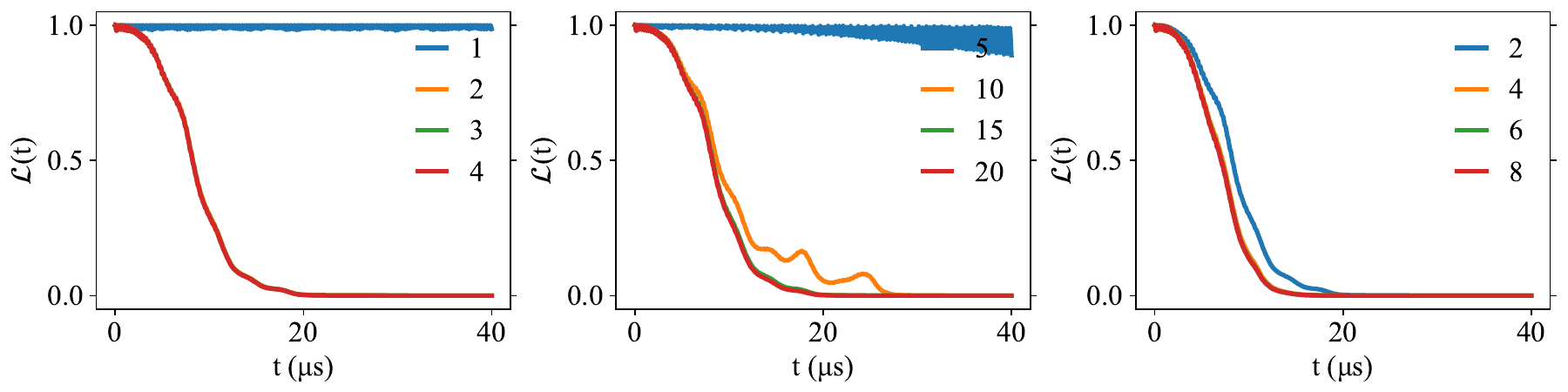}};
        \node[anchor=north west] at ([xshift=-9pt,yshift=5pt]image.north west) {(a)};
        \end{tikzpicture}
    \end{subfigure}    
    \hspace{1em}
    \begin{subfigure}[b]{0.3\linewidth}\label{fig:CCE_R_BATH}
        \centering
        \begin{tikzpicture}
        \node[anchor=south west, inner sep=0] (image) at (0,0)
        {\includegraphics[trim={10.05cm 0.0em 10.2cm 0.0em},clip,height=4.2cm,width=1.0\textwidth]{FIG_1_CONVERGENCE_CCE-eps-converted-to.pdf}};
        \node[anchor=north west] at ([xshift=-9pt,yshift=5pt]image.north west) {(b)};
        \end{tikzpicture}
    \end{subfigure}  
    \hspace{1em}
    \begin{subfigure}[b]{0.3\linewidth}\label{fig:CCE_R_DIPOLE}
        \centering
        \begin{tikzpicture}
        \node[anchor=south west, inner sep=0] (image) at (0,0)
        {\includegraphics[trim={20.15cm 0.0em 0.0em 0.0em},clip,height=4.2cm, width=1.0\textwidth]{FIG_1_CONVERGENCE_CCE-eps-converted-to.pdf}};
        \node[anchor=north west] at ([xshift=-9pt,yshift=5pt]image.north west) {(c)};
        \end{tikzpicture}
    \end{subfigure}   
    \caption{Correlation function $L(t$ computed as a function of the (a) order of the CCE calculations, (b) the radius of the bath (r$_{bath}$), and (c)  radius of the dipole-dipole interaction (r$_{dipole}$) for a single snapshot. The CCE order indicates the maximum size of the cluster, r$_{bath}$ defines the maximum distance from the central spin to the bath spin, and r$_{dipole}$ sets the maximum distance within which  two nuclear spins interact. Distances are given in \AA{}.}
    \label{fig:CCE}
\end{figure}

\section*{Quantum Model}

We carried out calculations with the quantum model for a representative amorphous ice sample (as defined in the main text), as a function of water density and graphite-water interaction, and the type of ions included in the solution.
We show below that in the absence of motional narrowing, the dependence of the computed coherence time T$_2$ on the surface-liquid interaction and the presence and concentration of ions is negligible. The dependence on density is trivial, stemming simply from a decreased number of protons at the interface, for lower densities of the liquid.

\subsection*{Simulations as a function of density} \label{QM_DENS}
 We compared results for the coherence time T2 obtained with the quantum model for two different densities of water:  $\sim$0.75 (see Fig.~\ref{fig:QM_DENSITY}) and 1 $grcm^{-3}$. For the lower density, we obtained T$_2$= 8.11 $\pm$ 0.71 $\mu$s, a value not surprisibly lower than the one obtained at the higher density (Fig. 2 in main text), since  the number of protons coupled to the qubit per unit volume decreases with density. The convergence parameters for the CCE expansion at lower density  were identical to those shown in Fig.~\ref{fig:CCE}.

\renewcommand{\thefigure}{S2}
\begin{figure}[!ht]
    \begin{subfigure}[b]{0.3\linewidth}\label{fig:QM_DENS_PROF}
        \centering
        \begin{tikzpicture}
        \node[anchor=south west, inner sep=0] (image) at (0,0)
        {\includegraphics[trim={0.0em 0.0em 0.0em 0.0em},clip,height = 4.2cm,width=1.0\textwidth]{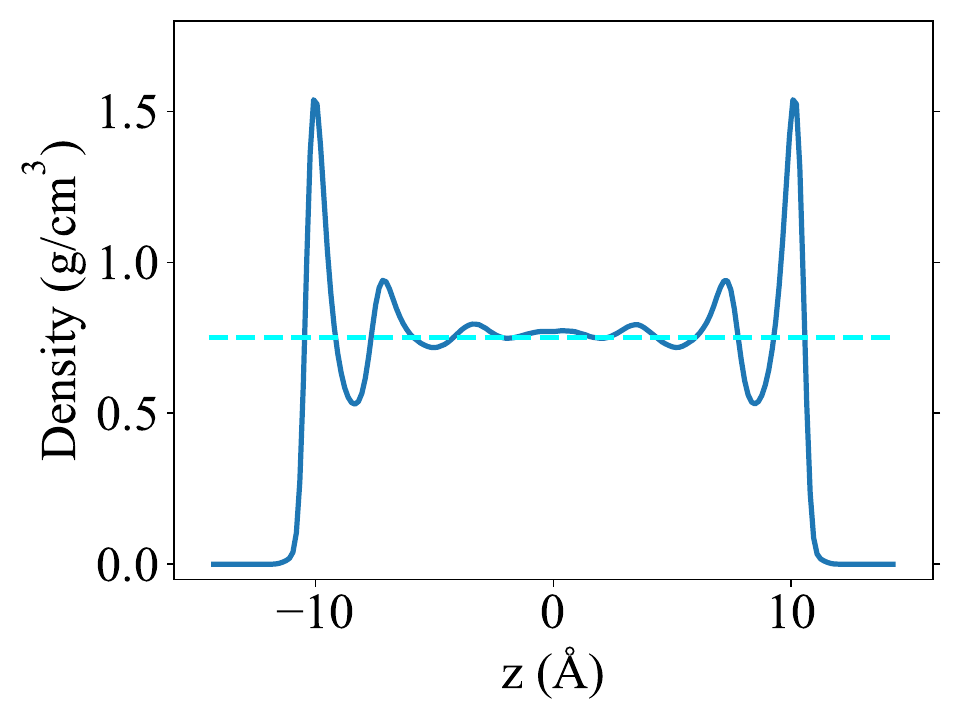}};
        \node[anchor=north west] at ([xshift=-9pt,yshift=5pt]image.north west) {(a)};
        \end{tikzpicture}
    \end{subfigure}    
    \hspace{3em}
    \begin{subfigure}[b]{0.3\linewidth}\label{fig:QM_DENS_T2}
        \centering
        \begin{tikzpicture}
        \node[anchor=south west, inner sep=0] (image) at (0,0)
        {\includegraphics[trim={0.0em 0.0em 0.0em 0.0em},clip,height = 4.2cm,width=1.0\textwidth]{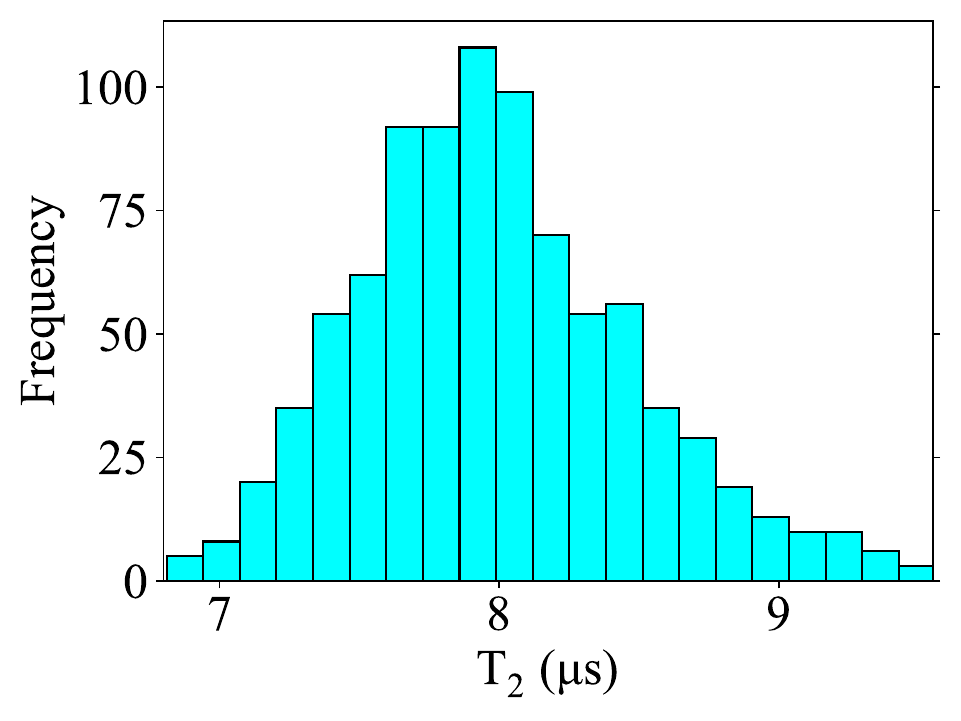}};
        \node[anchor=north west] at ([xshift=-9pt,yshift=5pt]image.north west) {(b)};
        \end{tikzpicture}
    \end{subfigure} 
    \caption{(a) Density of water along the $z$-direction perpendicular to the surface, with an average bulk density of $\sim$0.75 g/cm$^3$, represented by a dashed cyan line. (b) Distribution of computed coherence times for the liquid with density $\sim$0.75 g/cm$^3$; theaverage T$_2$ = 8.11 $\pm$ 0.71 $\mu$s.}
    \label{fig:QM_DENSITY}
\end{figure}

\subsection*{Strength of the surface-water interaction} \label{QM_INTERACTIONS}
Using the quantum model for a-H$_2$O, we carried out calculations  as a function of the LJ parameters for the carbon-oxygen interactions; the results are shown in Fig.~\ref{fig:QM_INTS}. We found that T$_2$ does not appreciably depend on the water-surface interaction,  since all computed T$_2$ values are within the standard deviation. This is in stark contrast with the results obtained with the semiclassical model, in the presence of motional narrowing, for the diffusing liquid at finite temperature.

\renewcommand{\thefigure}{S3}
\begin{figure}[!ht]
    \begin{subfigure}[b]{0.3\linewidth}\label{fig:QM_INT_DENS}
        \centering
        \begin{tikzpicture}
        \node[anchor=south west, inner sep=0] (image) at (0,0)
        {\includegraphics[trim={0.0em 0.0em 0.0em 0.0em},clip,height = 4.2cm,width=1.0\textwidth]{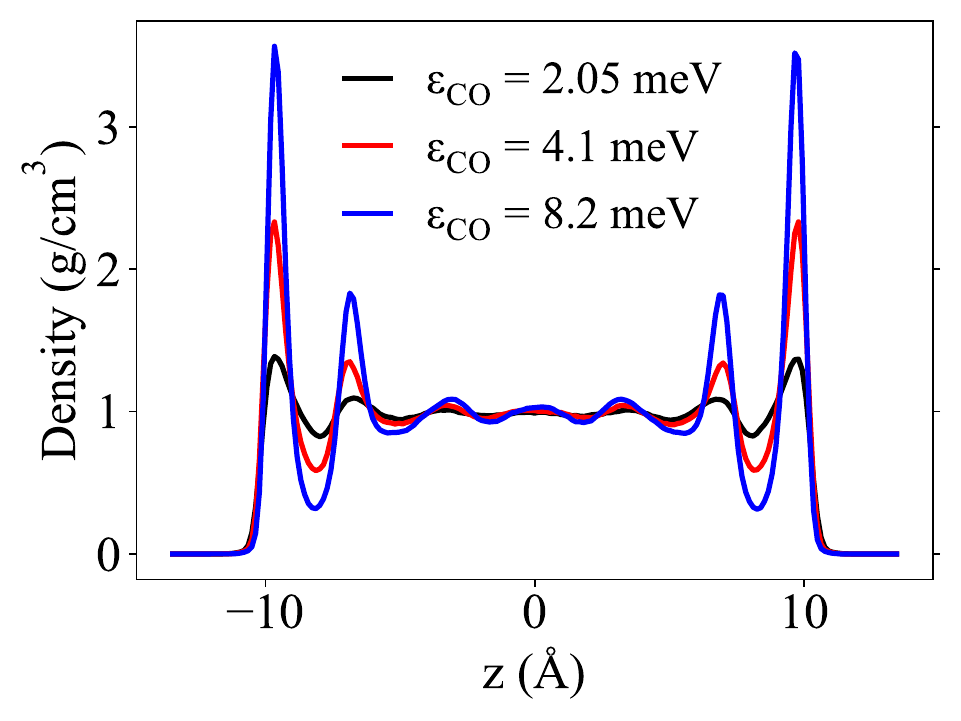}};
        \node[anchor=north west] at ([xshift=-9pt,yshift=5pt]image.north west) {(a)};
        \end{tikzpicture}
    \end{subfigure}  
    \hspace{3em}
    \begin{subfigure}[b]{0.3\linewidth}\label{fig:QM_INT_ALL}
        \centering
        \begin{tikzpicture}
        \node[anchor=south west, inner sep=0] (image) at (0,0)
        {\includegraphics[trim={0.0em 0.0em 0.0em 0.0em},clip,height = 4.2cm,width=1.0\textwidth]{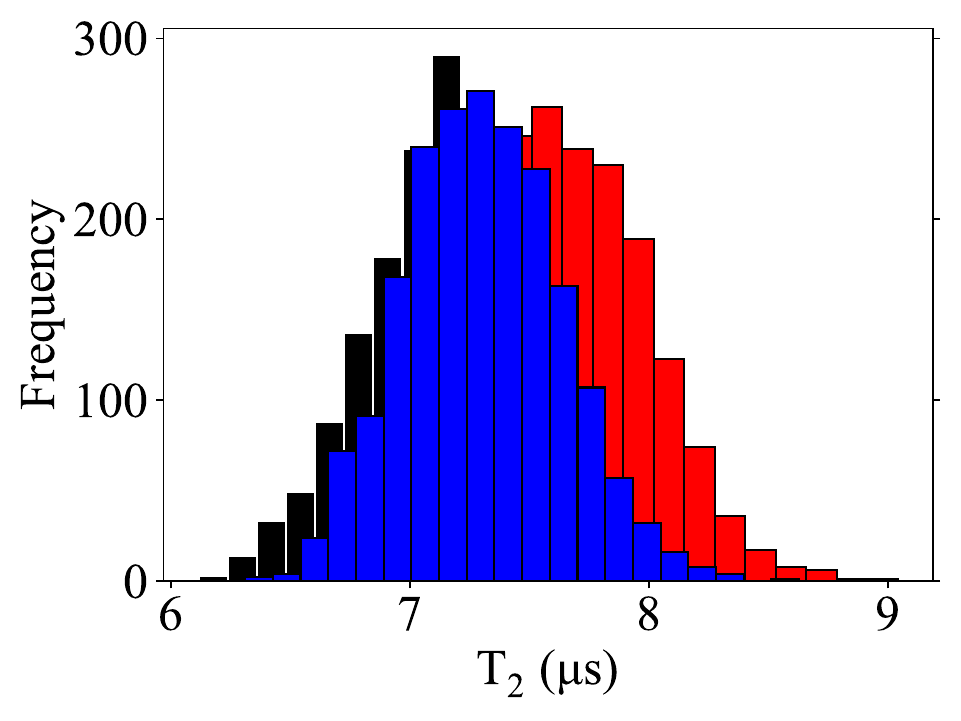}};
        \node[anchor=north west] at ([xshift=-9pt,yshift=5pt]image.north west) {(b)};
        \end{tikzpicture}
    \end{subfigure}
    \caption{(a) Density of water along the $z$-direction perpendicular to the surface for three different parameters defining the strength of the surface-water interaction. (b) Distribution of coherence times as a function of LJ interaction parameter $\epsilon_{CO}$: 2.05 meV (T$_2$ = 7.31 $\pm$ 0.32 $\mu$s),  4.1 meV (T$_2$ = 7.6 $\pm$ 0.38 $\mu$s), and  8.2 meV (T$_2$ = 7.21 $\pm$ 0.36 $\mu$s).}
    \label{fig:QM_INTS}
\end{figure}

\subsection*{Presence of ions} \label{QM_IONS}
We computed the coherence time of a-H$_2$O in the presence of ions, and the results are shown in Fig.~\ref{fig:T2_DIST_IONS}. Even though ions prefer to reside close to the surface (and hence the qubit)  their decoherence time is similar to that computed for pure water and largely independent on concentration.  The slight difference found at high concentrations of salts is mainly due to the difference in density at the interface discussed above.

\renewcommand{\thefigure}{S4}
\begin{figure}[!ht]
    \begin{subfigure}[b]{0.3\linewidth}\label{fig:T2_LICL}
        \centering
        \begin{tikzpicture}
        \node[anchor=south west, inner sep=0] (image) at (0,0)
        {\includegraphics[trim={0.0em 0.0em 0.0em 0.0em},clip,height = 4.2cm,width=1.0\textwidth]{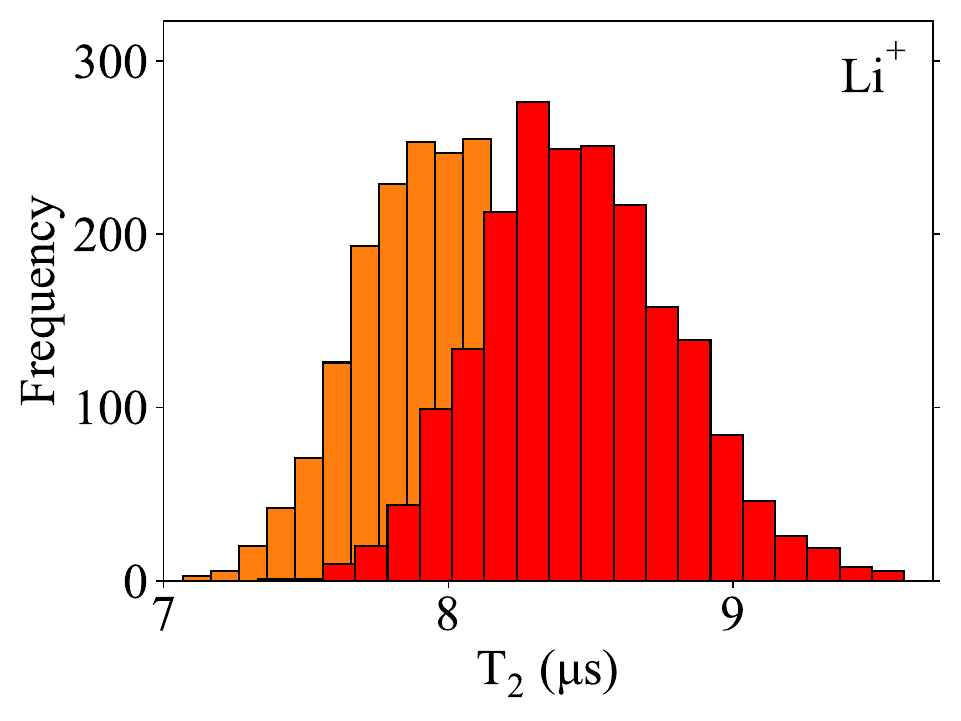}};
        \node[anchor=north west] at ([xshift=-9pt,yshift=5pt]image.north west) {(a)};
        \end{tikzpicture}
    \end{subfigure}
    \hspace{1em}
    \begin{subfigure}[b]{0.3\linewidth}\label{fig:T2_NACL}
        \centering
        \begin{tikzpicture}
        \node[anchor=south west, inner sep=0] (image) at (0,0)
        {\includegraphics[trim={0.0em 0.0em 0.0em 0.0em},clip,height = 4.2cm,width=1.0\textwidth]{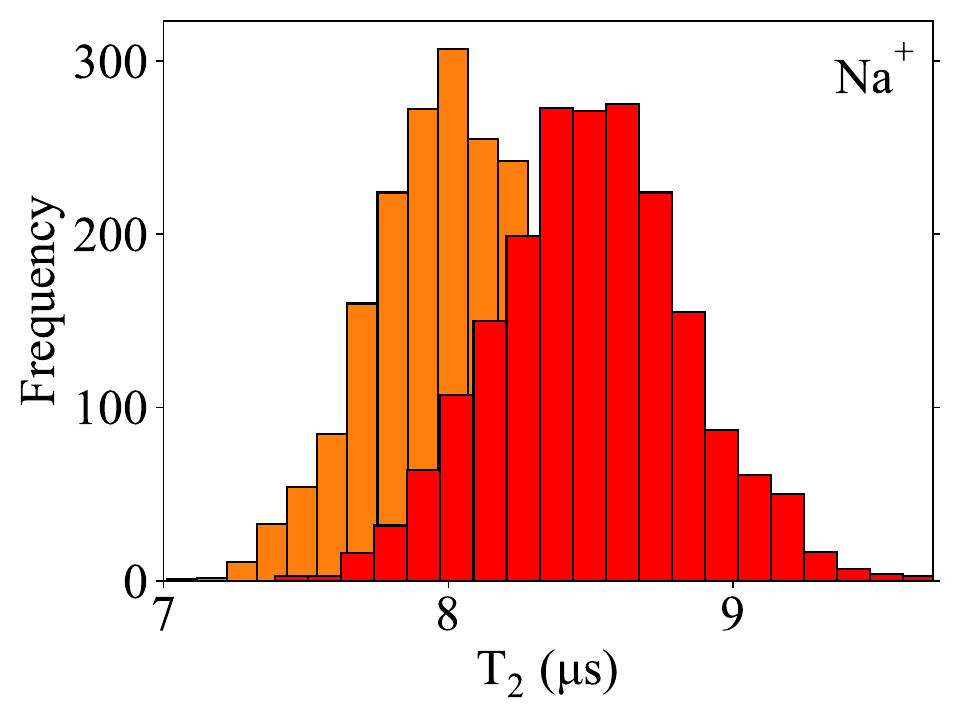}};
        \node[anchor=north west] at ([xshift=-9pt,yshift=5pt]image.north west) {(b)};
        \end{tikzpicture}
    \end{subfigure}
    \hspace{1em}
    \begin{subfigure}[b]{0.3\linewidth}\label{fig:T2_KCL}
        \centering
        \begin{tikzpicture}
        \node[anchor=south west, inner sep=0] (image) at (0,0)
        {\includegraphics[trim={0.0em 0.0em 0.0em 0.0em},clip,height = 4.2cm,width=1.0\textwidth]{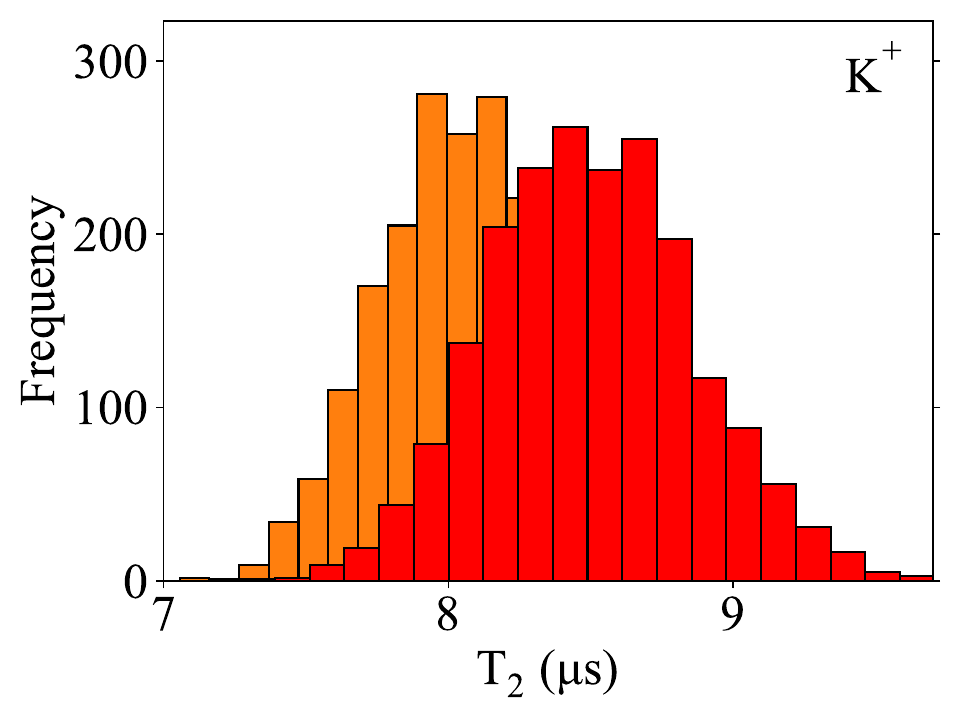}};
        \node[anchor=north west] at ([xshift=-9pt,yshift=5pt]image.north west) {(c)};
        \end{tikzpicture}
    \end{subfigure}
    \hspace{1em}
    \begin{subfigure}[b]{0.3\linewidth}\label{fig:T2_WAT}
        \centering
        \begin{tikzpicture}
        \node[anchor=south west, inner sep=0] (image) at (0,0)
        {\includegraphics[trim={0.0em 0.0em 0.0em 0.0em},clip,height = 4.2cm,width=1.0\textwidth]{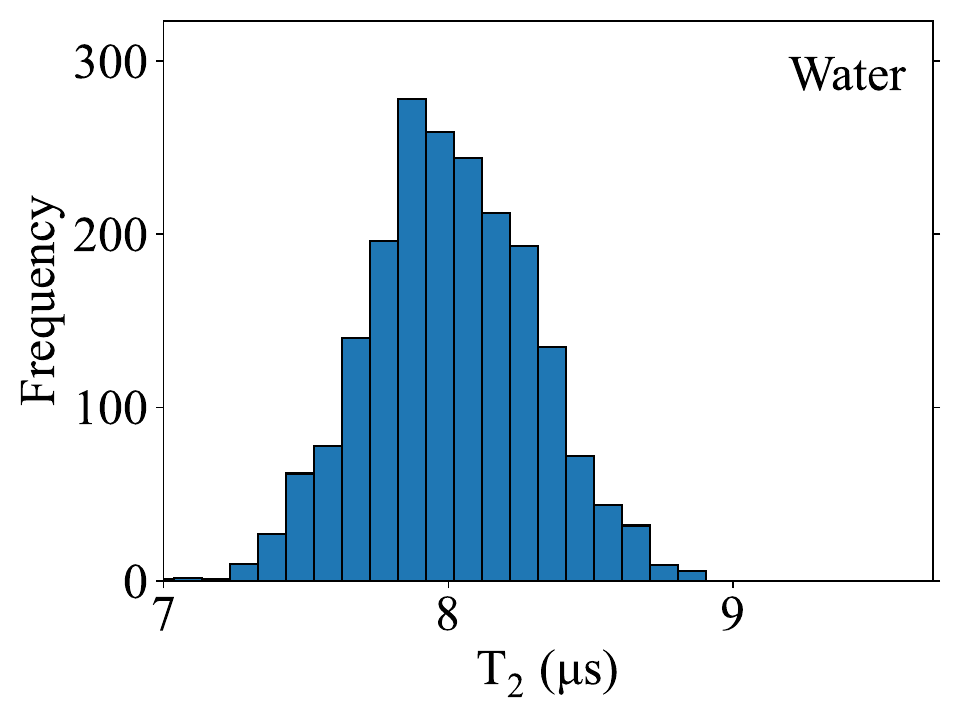}};
        \node[anchor=north west] at ([xshift=-9pt,yshift=5pt]image.north west) {(d)};
        \end{tikzpicture}
    \end{subfigure}  
    \hspace{1em}
    \begin{subfigure}[b]{0.3\linewidth}\label{fig:T2_MGCL}
        \centering
        \begin{tikzpicture}
        \node[anchor=south west, inner sep=0] (image) at (0,0)
        {\includegraphics[trim={0.0em 0.0em 0.0em 0.0em},clip,height = 4.2cm,width=1.0\textwidth]{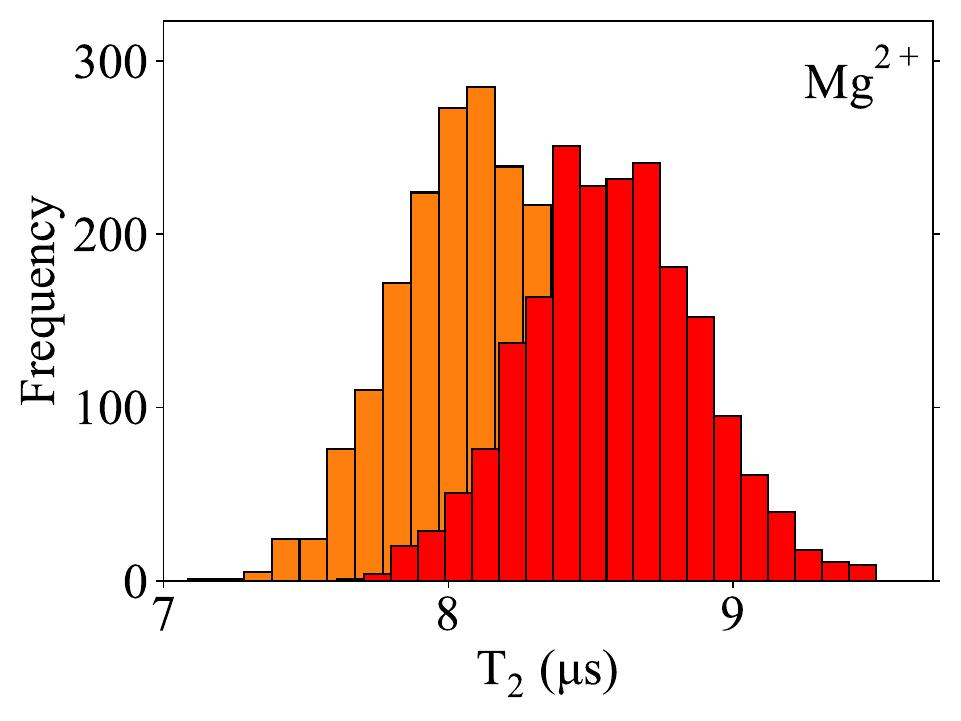}};
        \node[anchor=north west] at ([xshift=-9pt,yshift=5pt]image.north west) {(e)};
        \end{tikzpicture}
    \end{subfigure}  
    \hspace{1em}
    \begin{subfigure}[b]{0.3\linewidth}\label{fig:T2_CACL}
        \centering
        \begin{tikzpicture}
        \node[anchor=south west, inner sep=0] (image) at (0,0)
        {\includegraphics[trim={0.0em 0.0em 0.0em 0.0em},clip,height = 4.2cm,width=1.0\textwidth]{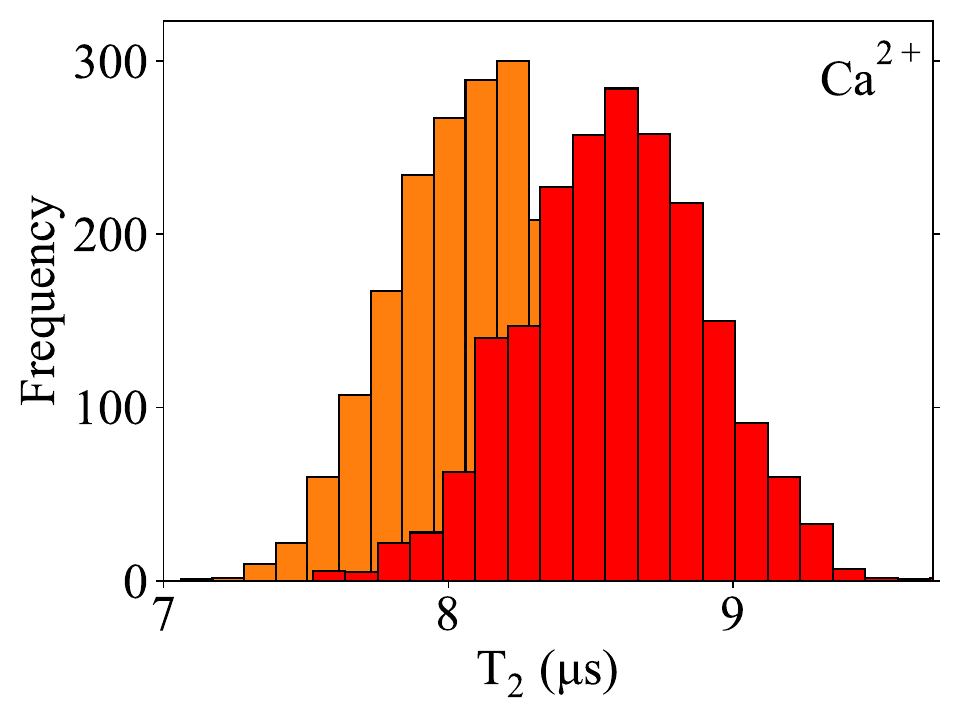}};
        \node[anchor=north west] at ([xshift=-9pt,yshift=5pt]image.north west) {(f)};
        \end{tikzpicture}
    \end{subfigure}  
    \caption{Distribution of coherence times computed with the quantum model for several ions solvated in water.  The results for pure  water, low, and high concentrations are shown in blue, orange, and red, respectively.}
    \label{fig:T2_DIST_IONS}
\end{figure}

\section*{Semiclassical Approximation} \label{SA}
\subsection*{Infinite time limit} \label{SA_INF}
Using only the $z$ component of the hyperfine interactions, 

\begin{equation} \label{eq20}
    A_{zz} = -  \frac{\mathcal{G}}{r^3} (3 \cos^2(\theta) - 1) 
\end{equation}

we can predict correlations in the infinite limit as:

\begin{equation} \label{eq21}
    C(\infty ) = \langle A_{zz} \rangle \langle A_{zz} \rangle
\end{equation}    

where the average $\langle A_{zz} \rangle$ is given by

\begin{equation} \label{eq22}
    \langle A_{zz} \rangle = - \mathcal{G} \langle\frac{1}{r^3}(3\cos^2(\theta) - 1)\rangle 
\end{equation}

\begin{equation} \label{eq23}
    \langle \frac{1}{r^{3}} (3 \cos^2(\theta) - 1) \rangle =  \frac{\int_{\frac{-L_{x}}{2}}^{\frac{L_{x}}{2}}  \,dx \int_{\frac{-L_{y}}{2}}^{\frac{L_{y}}{2}}  \,dy  \int_{0}^{L_{z}}  \,dz \frac{1}{r^{3}} (3 \cos^2(\theta) - 1) }{\int_{\frac{-L_{x}}{2}}^{\frac{L_{x}}{2}}  \,dx \int_{\frac{-L_{y}}{2}}^{\frac{L_{y}}{2}}  \,dy  \int_{0}^{L_{z}}  \,dz}
\end{equation}

and $\mathcal{G}$ = $\frac{\hbar \mu_0 \gamma_{e} \gamma_{proton}}{4\pi}$ = -79064.3 \AA$^3$ kHz. The distance $r$ in Cartesian coordinates and considering the qubit at the origin is:

\begin{equation} \label{eq24}
    r = \sqrt{x^2 + y^2 + z^2} 
\end{equation}

The angle $\theta$ is given by

\begin{equation} \label{eq25}
    \cos^2(\theta) = \left( \frac{\vec{B} \cdot \vec{r}}{|\vec{B}| |\vec{r}|} \right)^{2}
\end{equation}

where the quantization axis of the qubit, $\vec{B}$, is:

\begin{equation} \label{eq26}
    \vec{B} = (0, 0, L_z)
\end{equation}

The limits of integration allow us to consider the average distance between a fixed point (qubit at the origin) and a random point inside the simulation box (dimensions $L_{x}$, $L_{y}$, and $L_{z}$) with angle $\theta$ with respect to the $z$-axis. In order to consider a more realistic geometry, we subtracted the exclusion region ($\sim$2.5 \AA) in both limits of integration in the $z$ axis, since in this region there are no molecules of water due to the chosen solid-liquid interaction in the LJ potential.

\subsection*{Characterization of the nuclear spin noise} \label{SA_N}
To  compute time correlation functions C($\tau$) for the spin bath, we used snapshots extracted from  MD trajectories and the  PyCCE code\cite{OnizhukPyCCE:2021}.

\renewcommand{\thefigure}{S5}
\begin{figure}[!ht]
    \begin{subfigure}[b]{0.3\linewidth}\label{fig:CORR_LINEAR}
        \centering
        \begin{tikzpicture}
        \node[anchor=south west, inner sep=0] (image) at (0,0)
        {\includegraphics[trim={0.0em 0.0em 0.0em 0.0em},clip,height = 4.2cm,width=1.0\textwidth]{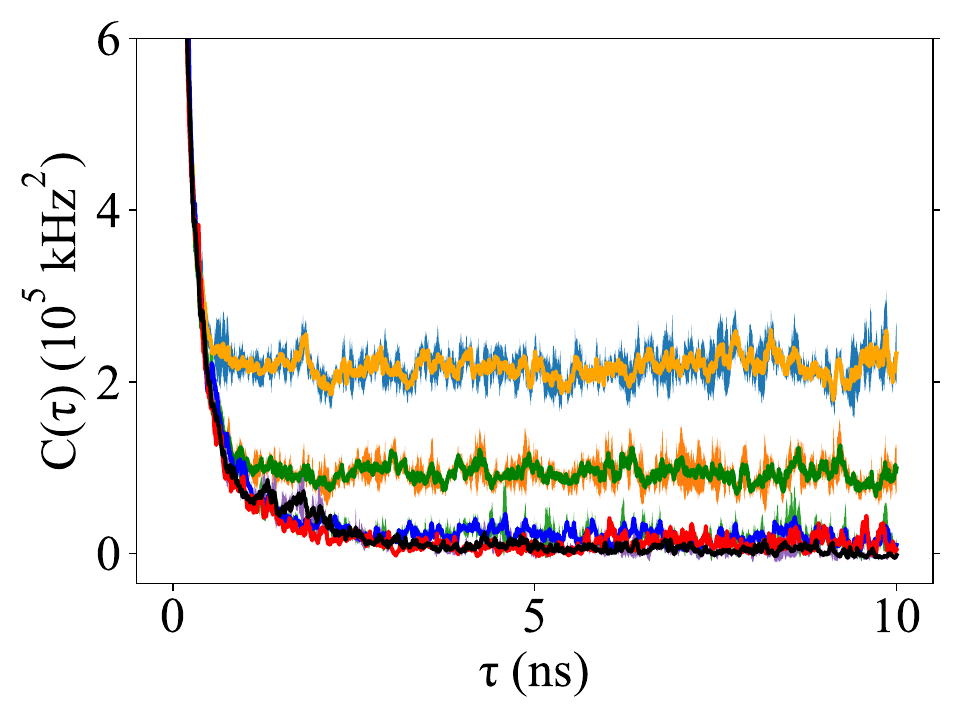}};
        \node[anchor=north west] at ([xshift=-9pt,yshift=5pt]image.north west) {(a)};
        \end{tikzpicture}
    \end{subfigure}
    \hspace{1em}
    \begin{subfigure}[b]{0.3\linewidth}\label{fig:CORRS_FIT}
        \centering
        \begin{tikzpicture}
        \node[anchor=south west, inner sep=0] (image) at (0,0)
        {\includegraphics[trim={0.0em 0.0em 0.0em 0.0em},clip,height = 4.2cm,width=1.0\textwidth]{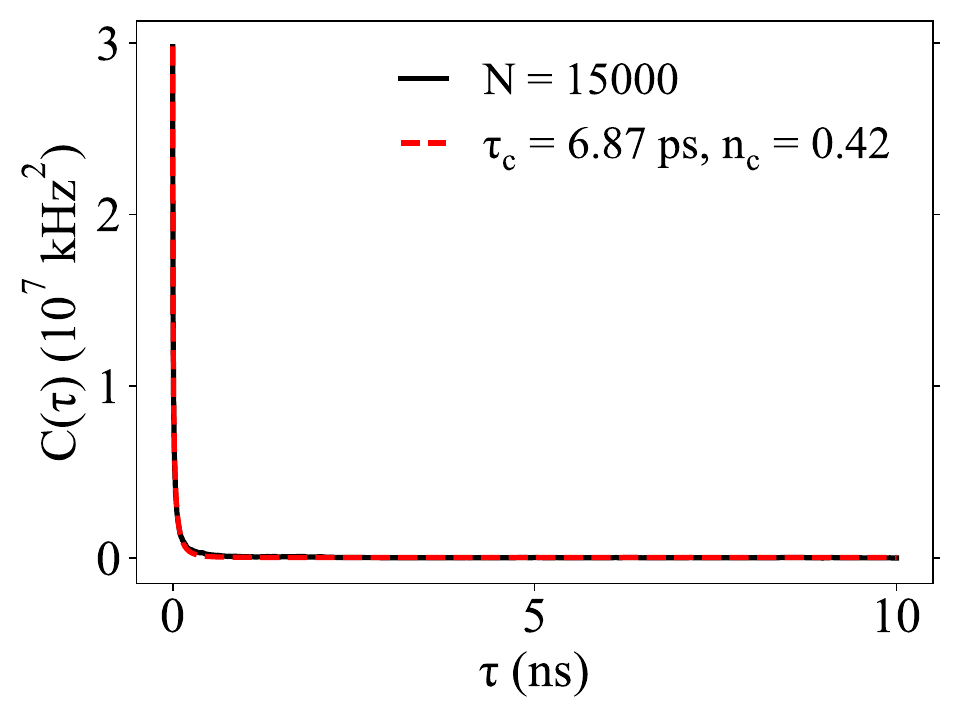}};
        \node[anchor=north west] at ([xshift=-9pt,yshift=5pt]image.north west) {(b)};
        \end{tikzpicture}
    \end{subfigure}  
    \hspace{1em}
    \begin{subfigure}[b]{0.3\linewidth}\label{fig:CORRS_INF_N}
        \centering
        \begin{tikzpicture}
        \node[anchor=south west, inner sep=0] (image) at (0,0)
        {\includegraphics[trim={0.0em 0.0em 0.0em 0.0em},clip,height = 4.2cm,width=1.0\textwidth]{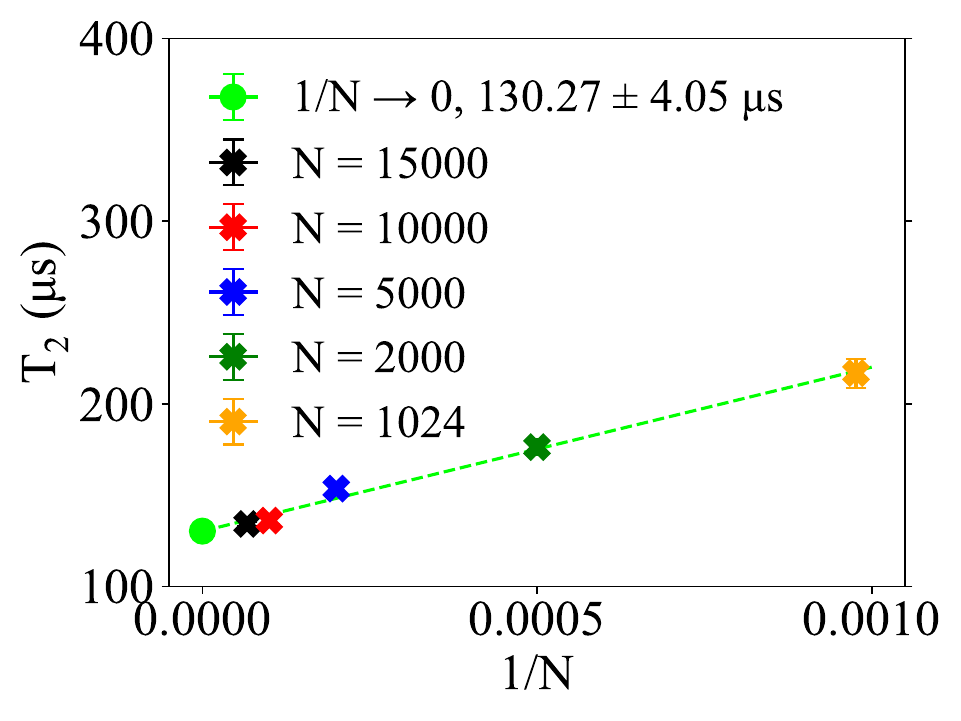}};
        \node[anchor=north west] at ([xshift=-9pt,yshift=5pt]image.north west) {(c)};
        \end{tikzpicture}
    \end{subfigure}  
    \caption{(a) Time autocorrelation function of protons as a function of the number of molecules of water included in the simulations cell: N = 1024, 2000, 5000, 10000 and 15000, represented by yellow, green, blue, red, and black solid lines, respectively. Shaded area represents standard deviation obtained from 3 different MD runs. (b) Time autocorrelation function obtained with N = 15000 and fitted to equation Eq. 7 in the main text, represented by a dashed red line. (c) Decoherence time T$_2$ as a function of 1/N with the extrapolated value for infinite size of the supercell shown as a green dot. }
    \label{fig:SEMI_APPROX_CORRS}
\end{figure}

We verified that computed time  correlation functions converge when at least 10,000 water molecules are included in the supercell.   The extrapolated values  to the infinite size supercell (1/N $\rightarrow$ 0) limit  are shown in Fig.~\ref{fig:SEMI_APPROX_CORRS}.  All results for T$_2$ shown in the main text and below are  obtained with 15,000 water molecule in the supercell.

\subsection*{Simulations as a function of density} \label{SA_DENS}
Results as a function of the density of water are shown in Fig. \ref{fig:SEMI_APPROX_CORRS_DENS} and  Table \ref{table:TIP4P_SPCE}. Within the statistical errors of our simulations, we find results for T$_2$ that are largely independent on density, in the infinite size limit.

\renewcommand{\thefigure}{S6}
\begin{figure}[!ht]
    \begin{subfigure}[b]{0.3\linewidth}\label{fig:CORRS_DENS_PROF}
        \centering
        \begin{tikzpicture}
        \node[anchor=south west, inner sep=0] (image) at (0,0)
        {\includegraphics[trim={0.0em 0.0em 0.0em 0.0em},clip,height = 4.2cm,width=1.0\textwidth]{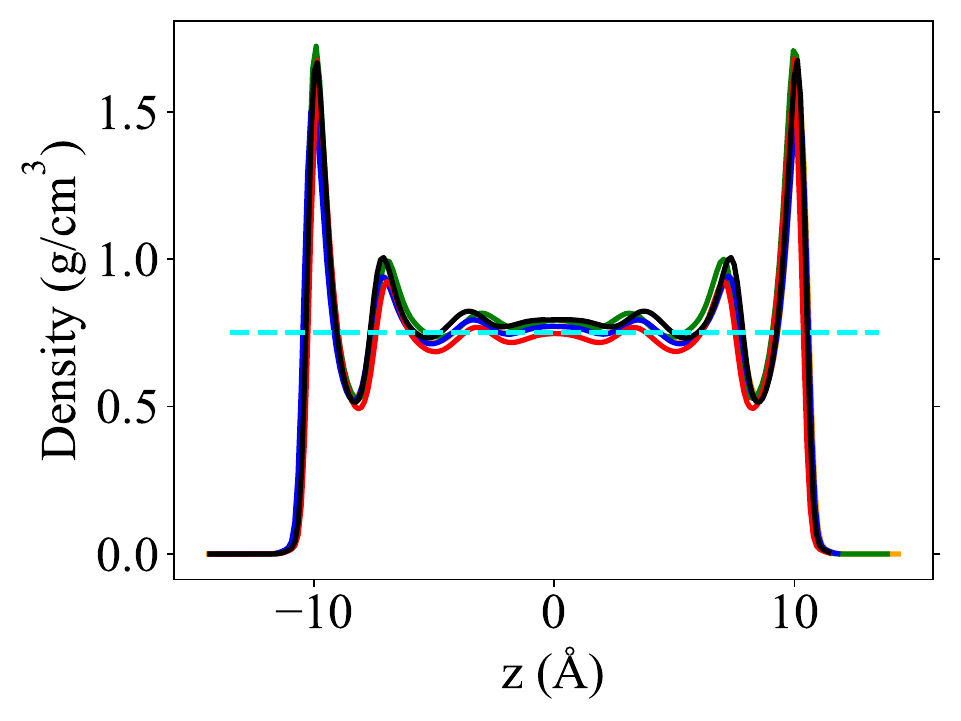}};
        \node[anchor=north west] at ([xshift=-9pt,yshift=5pt]image.north west) {(a)};
        \end{tikzpicture}
    \end{subfigure}
    \hspace{3em}
    \begin{subfigure}[b]{0.3\linewidth}\label{fig:CORRS_DENS_INF}
        \centering
        \begin{tikzpicture}
        \node[anchor=south west, inner sep=0] (image) at (0,0)
        {\includegraphics[trim={0.0em 0.0em 0.0em 0.0em},clip,height = 4.2cm,width=1.0\textwidth]{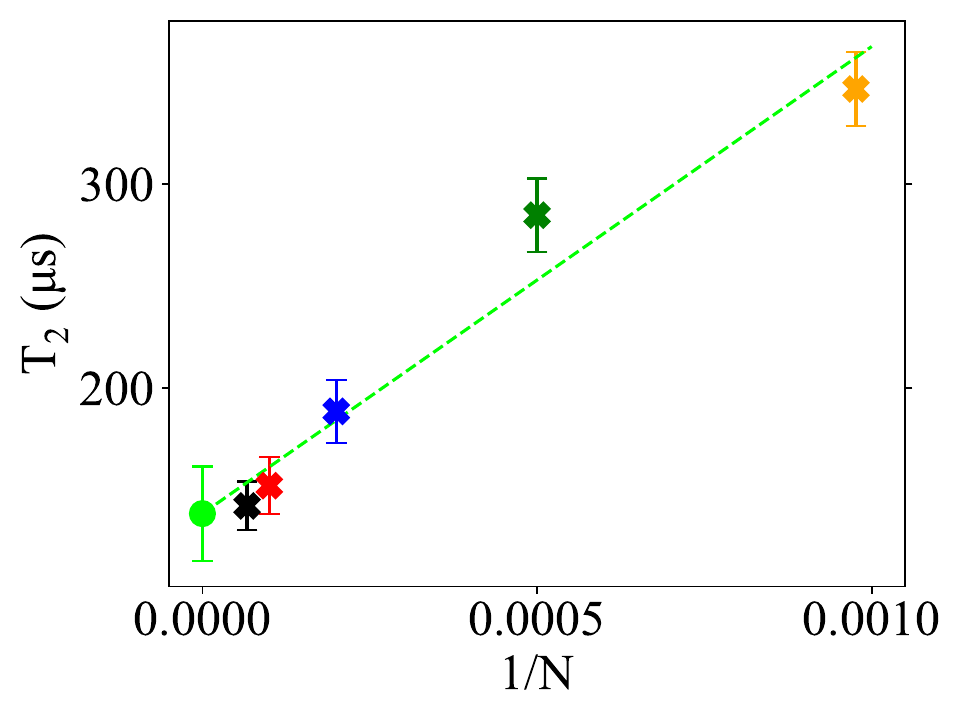}};
        \node[anchor=north west] at ([xshift=-9pt,yshift=5pt]image.north west) {(b)};
        \end{tikzpicture}
    \end{subfigure}  
    \caption{(a) Density along $z$-direction with bulk density $\sim$0.75 g/cm$^3$ represented by dashed cyan line. (b) Extrapolation of coherence times to the infinite size supercell with N = 1024, 2000, 5000, 10000 and 15000 represented by yellow, green, blue, red, and black solid lines, respectively.}
    \label{fig:SEMI_APPROX_CORRS_DENS}
\end{figure}

\subsection*{Comparison between force fields for water} \label{SA_FF}
 A comparison of the T$_2$ values computed with two different force fields for water is shown in Table \ref{table:TIP4P_SPCE}. The results obtained with TIP4P/2005 and  density of 1 gr$\times$cm$^{-3}$ show a difference of $\sim$15\% in the computed coherence time for the infinite size supercell with respect to the SPC/E force field. This difference stems from the fact that the SPC/E force field yields a larger diffusion coefficient of water compared to TIP4P FF. A faster diffusion  of molecules enhances the motional narrowing effect, dephasing the qubit over longer timescales.

\renewcommand{\thetable}{SI}
\begin{table}[!ht]
    \begin{center}
    \caption{Coherence times T$_2$ as a function of number (N) of molecules of water for two value of the densities of water ($\rho_{water}$, two different force fields (TIP4P/2005 and SPC/E). We also report the exponent $n$ obtained from the exponential fit of the correlation function $L(t)$.}
    \begin{tabular}{ccccccc}
    \hline
    \hline
    Water & \multicolumn{2}{c}{TIP4P/2005} & \multicolumn{2}{c}{TIP4P/2005} & \multicolumn{2}{c}{SPC/E}\\ [1 ex] 
    Molecules & \multicolumn{2}{c}{$\rho_{water}$ = 1.0 g/cm$^3$} & \multicolumn{2}{c}{$\rho_{water} \sim$ 0.75 g/cm$^3$} & \multicolumn{2}{c}{$\rho_{water}$ = 1.0 g/cm$^3$}\\
    N & T$_2$, $\mu$s & $n$ & T$_2$, $\mu$s & $n$ & T$_2$, $\mu$s & $n$ \\
    \hline
    1024 & 216.73 $\pm$ 7.97 & 0.99 & 346.61$\pm$ 18.0 & 0.99 & 283.36 $\pm$ 11.02 & 0.99 \\
    2000 & 176.29 $\pm$ 4.17 & 0.99 & 284.75 $\pm$ 18.0 & 0.99 & 222.05 $\pm$ 6.09 & 0.99\\
    5000 & 153.57 $\pm$ 2.63 & 0.99 & 188.63 $\pm$ 15.36 & 0.99 & 180.29 $\pm$ 3.18 & 0.99\\
    10000 & 136.52 $\pm$ 2.47 & 1.0 & 152.2 $\pm$ 13.85 & 1.0 & 157.51 $\pm$ 2.83 & 1.0\\
    15000 & 134.19 $\pm$ 1.32 & 1.0 & 142.24 $\pm$ 12.27 & 1.0 & 152.61 $\pm$ 2.46 & 1.0\\
    INF & 130.27 $\pm$ 4.05 & - & 138.49 $\pm$ 23.07 & - & 146.52 $\pm$ 5.14 & -\\
    \hline
    \hline
    \end{tabular}
    \label{table:TIP4P_SPCE}
    \end{center}
\end{table}

\subsection*{Strength of the surface-water interaction} \label{SA_HYDRO}
We analyzed the reorientation dynamics of water molecules in the  5 regions of the cell defined in Fig.~\ref{fig:SEMI_APPROX_EPS} (a) by computing several characteristic time scales.

\renewcommand{\thefigure}{S7}
\begin{figure}[!ht]
    \begin{subfigure}[b]{0.3\linewidth}\label{fig:EPS_DENS_PROF}
        \centering
        \begin{tikzpicture}
        \node[anchor=south west, inner sep=0] (image) at (0,0)
        {\includegraphics[trim={0.0em 0.0em 0.0em 0.0em},clip,height = 4.2cm,width=1.0\textwidth]{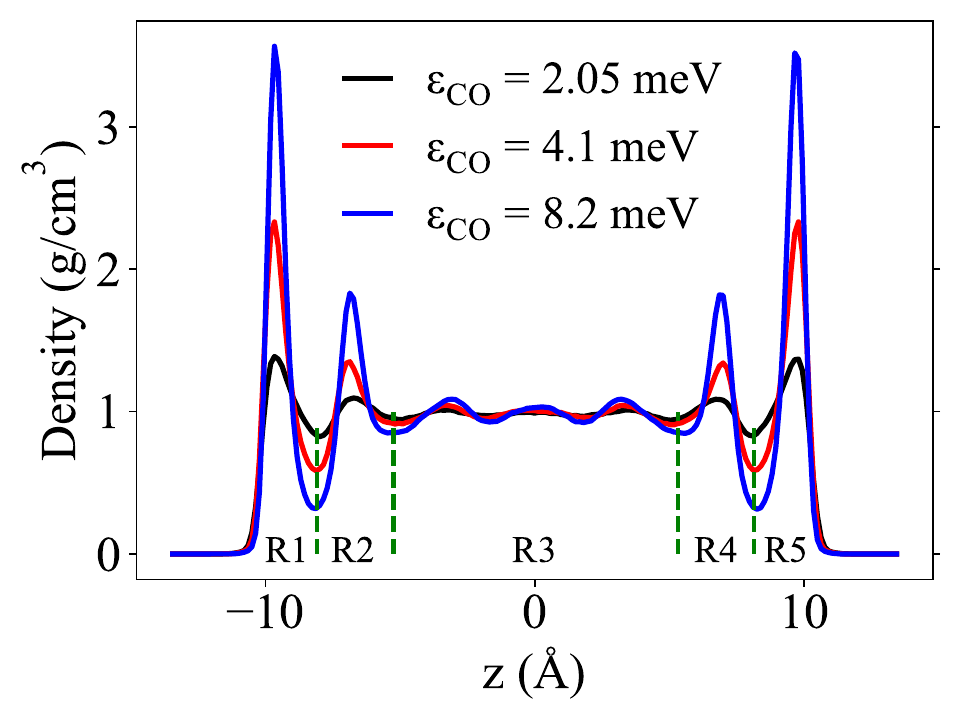}};
        \node[anchor=north west] at ([xshift=-9pt,yshift=5pt]image.north west) {(a)};
        \end{tikzpicture}
    \end{subfigure}
    \hspace{3em}
    \begin{subfigure}[b]{0.3\linewidth}\label{fig:EPS_NHB}
        \centering
        \begin{tikzpicture}
        \node[anchor=south west, inner sep=0] (image) at (0,0)
        {\includegraphics[trim={0.0em 0.0em 0.0em 0.0em},clip,height = 4.2cm,width=1.0\textwidth]{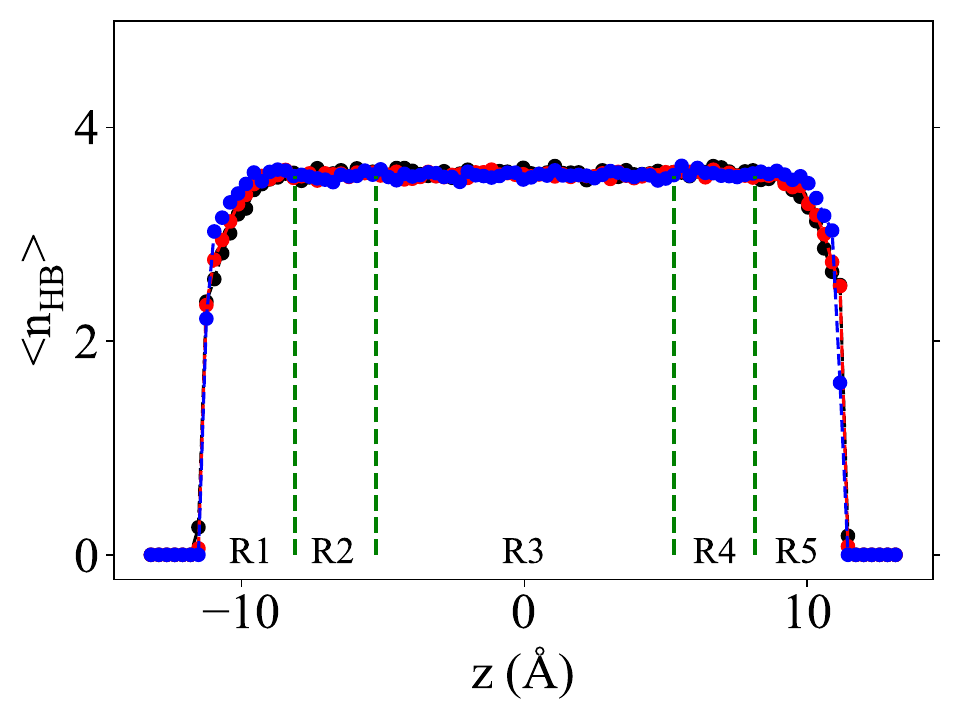}};
        \node[anchor=north west] at ([xshift=-9pt,yshift=5pt]image.north west) {(b)};
        \end{tikzpicture}
    \end{subfigure}  
    \caption{(a) Density distribution and (b) average number of hydrogen bonds along the $z$-axis perpendicular to the surface. We defined three different regions delimited by dashed green lines: an interfacial region (R1, R5) corresponding to distances over which the density of the liquid is significantly different from that of bulk water; a central region (R3) where the density of the liquid is the same as in the bulk; and an intermediate region (R2, R4) with smaller density variations than in the interfacial  regions.}
    \label{fig:SEMI_APPROX_EPS}
\end{figure}

\subsubsection*{Reorientation times}
The reorientation correlation function is defined as follows:

\begin{equation} \label{eq27}
    RCF(\tau) = \frac{1}{N_w}\langle \sum_{i=1}^{N_w}\mu(\tau)\cdot\mu(0) \rangle
\end{equation}

where N$_w$ is the number of water molecules in a given region of the cell and $\mu(\tau)$ is the water dipole moment at time $\tau$. This time correlation is well described by a stretch exponential function:

\begin{equation} \label{eq28}
    RCF(\tau) = exp[-(\frac{\tau}{\tau_{DM}})^{\beta}]
\end{equation}

where $\beta$ is the stretch exponent and $\tau_{DM}$ is the characteristic reorientation time. The results for all different regions of the system are shown in Table \ref{table:TAU_DM_EPSILONS}.

\renewcommand{\thetable}{SII}
\begin{table}[!ht]
    \begin{center}
    \caption{Reorientation times of water dipole moment for all regions of the liquid as defined in Fig.~\ref{fig:SEMI_APPROX_EPS},  as a function of the Lennard-Jones (LJ) parameter $\epsilon_{CO}$. Uncertainties in $\tau_{DM}$ are given in the last digits in parenthesis. We also report values of the stretch exponent $\beta$.}
    \centering
    \begin{tabular}{ccccccc}
    \hline
    \hline
    Region & \multicolumn{2}{c}{$\epsilon_{CO}$ = 2.05 meV} & \multicolumn{2}{c}{$\epsilon_{CO}$ = 4.1 meV} & \multicolumn{2}{c}{$\epsilon_{CO}$ = 8.2 meV} \\ [3pt]
     & $\tau_{DM}$, ps & $\beta$ & $\tau_{DM}$, ps & $\beta$ & $\tau_{DM}$, ps & $\beta$ \\ [3pt]
    \hline
    \rowcolor{Gray}
R1 & 4.47(8) & 0.78(1) & 4.62(6) & 0.78(1) & 5.24(3) & 0.75(1) \\
R2 & 4.47(9) & 0.78(1) & 4.56(7) & 0.77(1) & 4.97(1) & 0.75(1) \\
\rowcolor{Gray}
R3 & 4.46(9) & 0.78(1) & 4.49(4) & 0.78(1) & 4.86(4) & 0.76(1) \\
R4 & 4.45(9) & 0.77(1) & 4.55(6) & 0.77(1) & 4.98(6) & 0.76(1) \\
\rowcolor{Gray}
R5 & 4.43(7) & 0.78(1) & 4.59(3) & 0.77(1) & 5.27(1) & 0.74(1) \\
    \hline
    \hline
    \end{tabular}
    \label{table:TAU_DM_EPSILONS}
    \end{center}
\end{table}

\subsubsection*{Average number of hydrogen bonds}
We computed the average number of hydrogen bonds along the $z$ axis perpendicular to the surface, defined as the distance between two oxygen atoms r$_{OO}$ $\leq$ 3.5 \AA{} and the angle formed by the oxygen acceptor-oxygen donor-hydrogen donor $\theta$ $\leq$ 30$^{\circ}$ \cite{Antipova:2013}. The results for different values of the eLJ parameter $\epsilon_{CO}$ are shown in Fig. \ref{fig:SEMI_APPROX_EPS}(b). 

\subsubsection*{Average lifetime of hydrogen bonds}
The continuous average of the hydrogen bond lifetime can be evaluated through the following correlation function \cite{Rapaport:1983}:

\begin{equation} \label{eq29}
    C_{HB}(\tau) = \langle \frac{\sum_{ij}s_{ij}(\tau) \cdot s_{ij}(0)}{\sum_{ij}s^2_{ij}(0)} \rangle
\end{equation}

where the value of $s_{ij}$ can only have a single transition from 1 to 0 when the bond is first broken, but cannot  return to the value of 1 if the same bond is reformed. C$_{HB}(\tau)$ is calculated directly from the MD trajectories and fitted to a bi-exponential function:

\begin{equation} \label{eq30}
    C_{HB}(\tau) = A_1 exp(\frac{-\tau}{\tau_1}) + A_2 exp(\frac{-\tau}{\tau_2})
\end{equation}

where the average lifetime of hydrogen bonds $\tau_{HB}$ is given by the product $A_1 \tau_1 + A_2 \tau_2$. Results for different values of $\epsilon_{CO}$ are shown in Table \ref{table:TAU_HB_EPSILONS}.

\renewcommand{\thetable}{SIII}
\begin{table}[!ht]
    \begin{center}
    \caption{Average lifetime of hydrogen bonds in all regions of the liquid, as defined in Fig.~\ref{fig:SEMI_APPROX_EPS}, as a function of the Lennard-Jones (LJ) parameter $\epsilon_{CO}$. Uncertainties of $\tau_{HB}$ are given in the last digits in parenthesis.}
    \begin{tabular}{cccc}
    \hline
    \hline
    Region & $\epsilon_{CO}$ = 2.05 meV & $\epsilon_{CO}$ = 4.1 meV & $\epsilon_{CO}$ = 8.2 meV \\ [3pt]
     & $\tau_{HB}$, ps & $\tau_{HB}$, ps & $\tau_{HB}$, ps \\ [3pt]
    \hline
    \rowcolor{Gray}
R1 & 0.73(4) & 0.69(5) & 0.69(5) \\
R2 & 0.74(4) & 0.68(3) & 0.72(4) \\
\rowcolor{Gray}
R3 & 0.78(1) & 0.76(2) & 0.76(4) \\
R4 & 0.78(4) & 0.75(3) & 0.68(4) \\
\rowcolor{Gray}
R5 & 0.70(2) & 0.73(1) & 0.70(4) \\
    \hline
    \hline
    \end{tabular}
    \label{table:TAU_HB_EPSILONS}
    \end{center}
\end{table}

\subsubsection*{Reorientation times of OH bonds}
If we consider the vector formed by the OH bond of each water molecule in Eq.~\ref{eq27} we obtain the reorientation time of such bond. The results for different values of $\epsilon_{CO}$ are shown in Table \ref{table:TAU_OH_EPSILONS}. 

\renewcommand{\thetable}{SIV}
\begin{table}[!ht]
    \begin{center}
    \caption{Reorientation times of OH bonds for all regions of the liquid, as a function of the Lennard-Jones (LJ) parameter $\epsilon_{CO}$. Uncertainties in $\tau_{OH}$ are given in the last digits in parenthesis.}
    \centering
    \begin{tabular}{ccccccc}
    \hline
    \hline
    Region & \multicolumn{2}{c}{$\epsilon_{CO}$ = 2.05 meV} & \multicolumn{2}{c}{$\epsilon_{CO}$ = 4.1 meV} & \multicolumn{2}{c}{$\epsilon_{CO}$ = 8.2 meV} \\ [3pt]
     & $\tau_{OH}$, ps & $\beta$ & $\tau_{OH}$, ps & $\beta$ & $\tau_{OH}$, ps & $\beta$ \\ [3pt]
    \hline
    \rowcolor{Gray}
R1 & 4.74(1) & 0.82(1) & 4.84(4) & 0.82(1) & 5.32(2) & 0.81(1) \\
R2 & 4.77(9) & 0.82(1) & 4.82(7) & 0.82(1) & 5.21(2) & 0.81(1) \\
\rowcolor{Gray}
R3 & 4.81(9) & 0.83(1) & 4.83(5) & 0.83(1) & 5.16(4) & 0.82(1) \\
R4 & 4.77(9) & 0.82(1) & 4.83(6) & 0.83(1) & 5.21(3) & 0.82(1) \\
\rowcolor{Gray}
R5 & 4.71(8) & 0.82(1) & 4.81(4) & 0.82(2) & 5.36(4) & 0.81(1) \\
    \hline
    \hline
    \end{tabular}
    \label{table:TAU_OH_EPSILONS}
    \end{center}
\end{table}

\subsection*{Effect of confinement} \label{SA_CONF}
In order to explore the effect of confinement on our calculations of coherence times, we carried out MD simulations with two additional separation distances between graphene layers: $\sim$15 \AA{} and $\sim$52 \AA{}. The extrapolation of coherence times to the infinite size supercell is shown in Fig.~\ref{fig:SA_INF_CORSS} (a) and (b). We note that the results obtained with the largest supercell (N=15,000) yield similar values of T$_2$ for all distances,  This supports our previous findings, that the qubit is mostly sensitive to the structure and dynamics of water in close proximity of the solid-liquid interface.

 A comparison of the analytical form of the  correlation function in the infinite time limit and that computed from MD trajectories is shown in Figs.~\ref{fig:SA_INF_CORSS} (c)-(e), showing  good agreement between the two approaches. 

 

\renewcommand{\thefigure}{S8}
\begin{figure}[!ht]
    \begin{subfigure}[b]{0.3\linewidth}\label{fig:INF_N_15A}
        \centering
        \begin{tikzpicture}
        \node[anchor=south west, inner sep=0] (image) at (0,0)
        {\includegraphics[trim={0.0em 0.0em 0.0em 0.0em},clip,height = 4.2cm,width=1.0\textwidth]{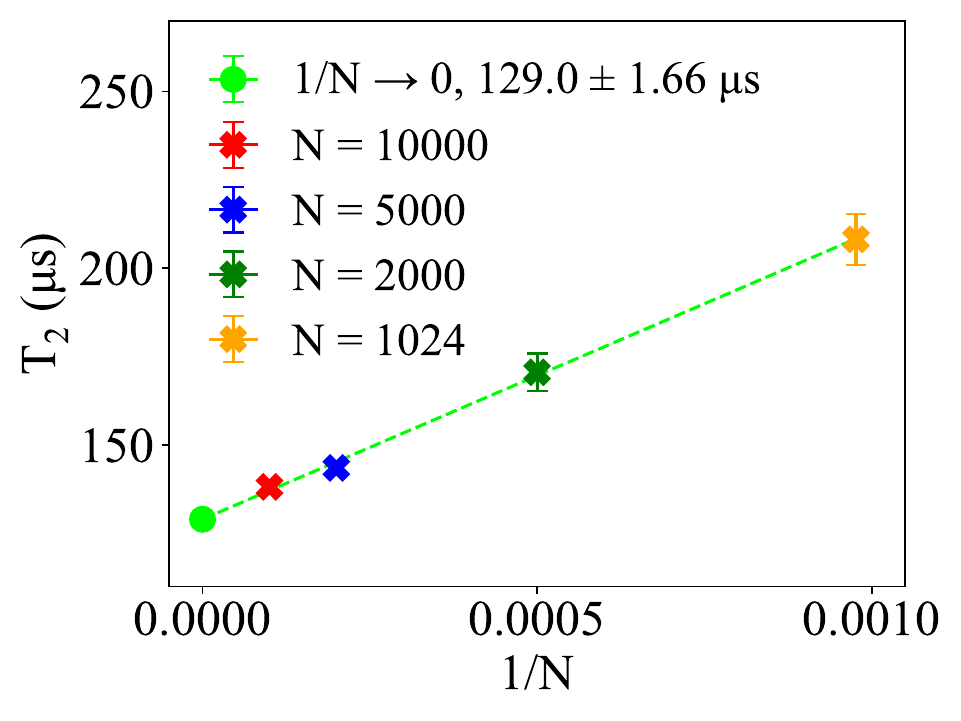}};
        \node[anchor=north west] at ([xshift=-9pt,yshift=5pt]image.north west) {(a)};
        \end{tikzpicture}
    \end{subfigure}
    \hspace{1em}
    \begin{subfigure}[b]{0.3\linewidth}\label{fig:INF_N_52A}
        \centering
        \begin{tikzpicture}
        \node[anchor=south west, inner sep=0] (image) at (0,0)
        {\includegraphics[trim={0.0em 0.0em 0.0em 0.0em},clip,height = 4.2cm,width=1.0\textwidth]{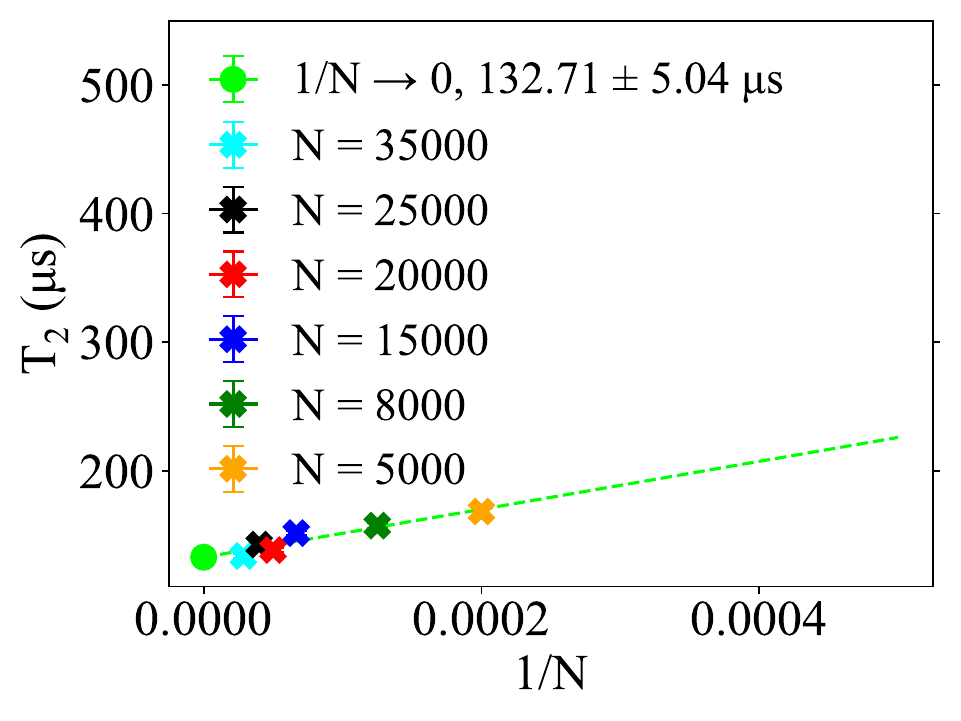}};
        \node[anchor=north west] at ([xshift=-9pt,yshift=5pt]image.north west) {(b)};
        \end{tikzpicture}
    \end{subfigure}  
    \hspace{1em}
    \begin{subfigure}[b]{0.3\linewidth}\label{fig:INF_15A}
        \centering
        \begin{tikzpicture}
        \node[anchor=south west, inner sep=0] (image) at (0,0)
        {\includegraphics[trim={0.0em 0.0em 0.0em 0.0em},clip,height = 4.2cm,width=1.0\textwidth]{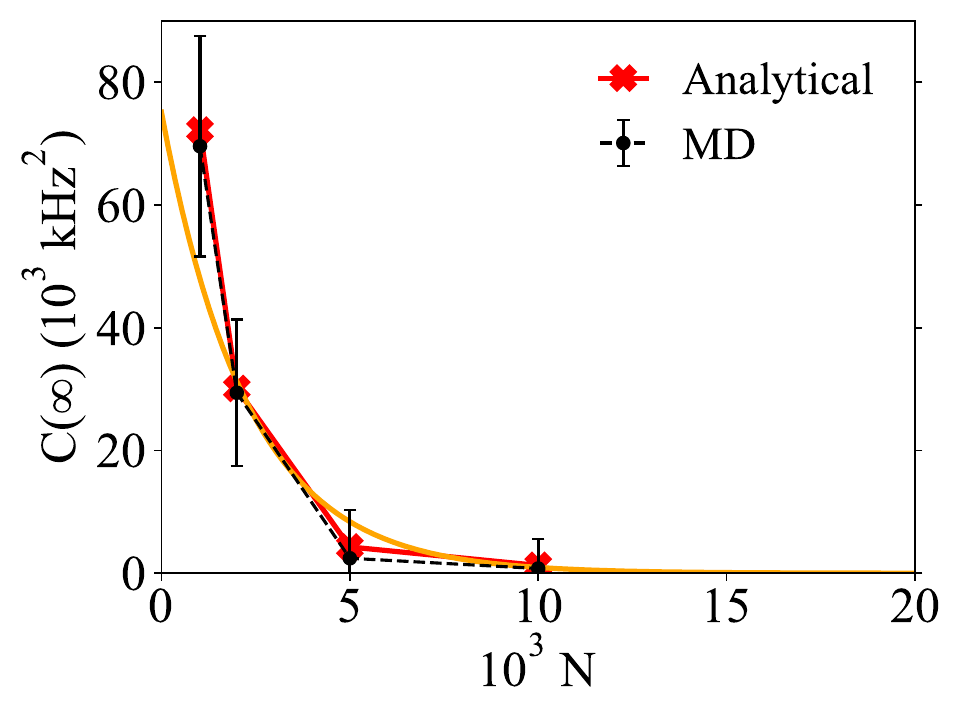}};
        \node[anchor=north west] at ([xshift=-9pt,yshift=5pt]image.north west) {(c)};
        \end{tikzpicture}
    \end{subfigure}
    \hspace{1em}
    \begin{subfigure}[b]{0.3\linewidth}\label{fig:INF_25A}
        \centering
        \begin{tikzpicture}
        \node[anchor=south west, inner sep=0] (image) at (0,0)
        {\includegraphics[trim={0.0em 0.0em 0.0em 0.0em},clip,height = 4.2cm,width=1.0\textwidth]{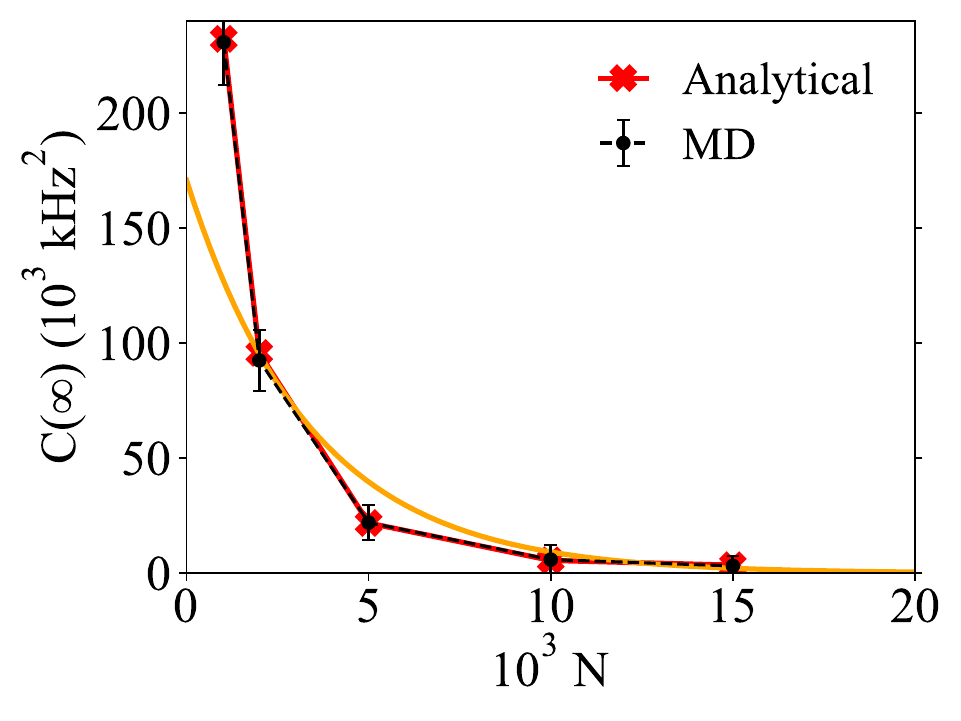}};
        \node[anchor=north west] at ([xshift=-9pt,yshift=5pt]image.north west) {(d)};
        \end{tikzpicture}
    \end{subfigure}  
    \hspace{1em}
    \begin{subfigure}[b]{0.3\linewidth}\label{fig:INF_52A}
        \centering
        \begin{tikzpicture}
        \node[anchor=south west, inner sep=0] (image) at (0,0)
        {\includegraphics[trim={0.0em 0.0em 0.0em 0.0em},clip,height = 4.2cm,width=1.0\textwidth]{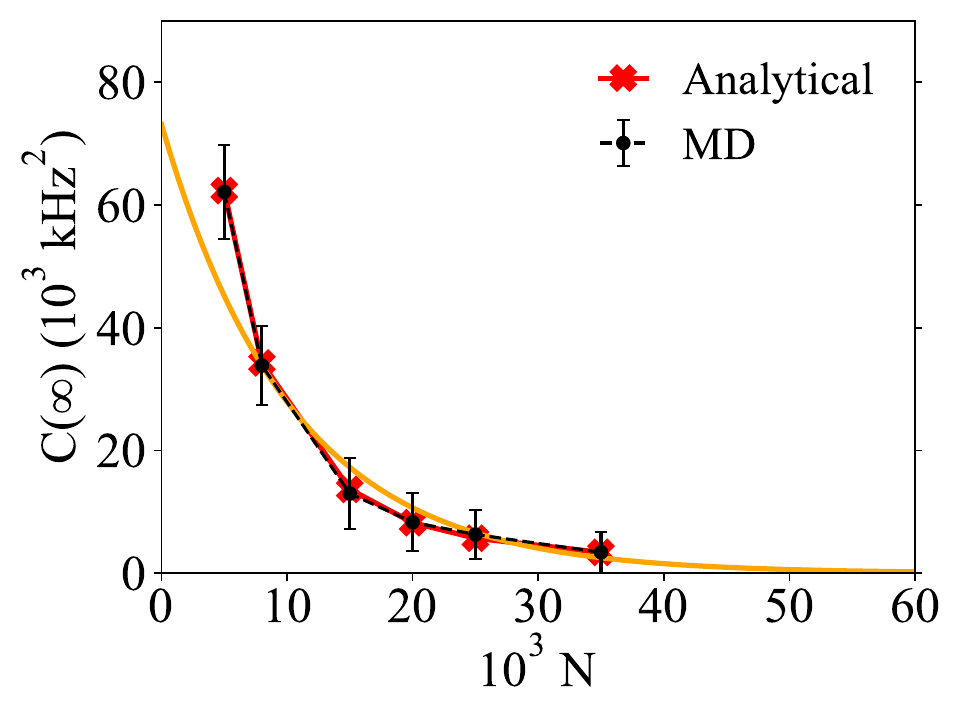}};
        \node[anchor=north west] at ([xshift=-9pt,yshift=5pt]image.north west) {(e)};
        \end{tikzpicture}
    \end{subfigure}  
    \caption{Extrapolation of T$_2$ to the infinite supercell size limit for h$_{conf}$ equal to (a) $\sim$15 \AA{}, and (b) $\sim$52 \AA{}. Time correlations in the infinite time limit for (c) 15 \AA, (d) 28 \AA, and (e) 52 \AA.} 
    \label{fig:SA_INF_CORSS}
\end{figure}

\subsection*{Presence of ions} \label{SA_IONS}
The reorientation times of the water dipole moment for all different regions of the liquid with low and high salt concentrations are shown in Tables \ref{table:SA_TAU_DM_LOW} and \ref{table:SA_TAU_DM_MONO_HIGH}, \ref{table:SA_TAU_DM_DI_HIGH}, respectively.

\renewcommand{\thefigure}{S9}
\begin{figure}[!ht]
    \begin{subfigure}[b]{0.3\linewidth}\label{fig:IONS_DENS}
        \centering
        \begin{tikzpicture}
        \node[anchor=south west, inner sep=0] (image) at (0,0)
        {\includegraphics[trim={0.0em 0.0em 0.0em 0.0em},clip,height = 4.2cm,width=1.0\textwidth]{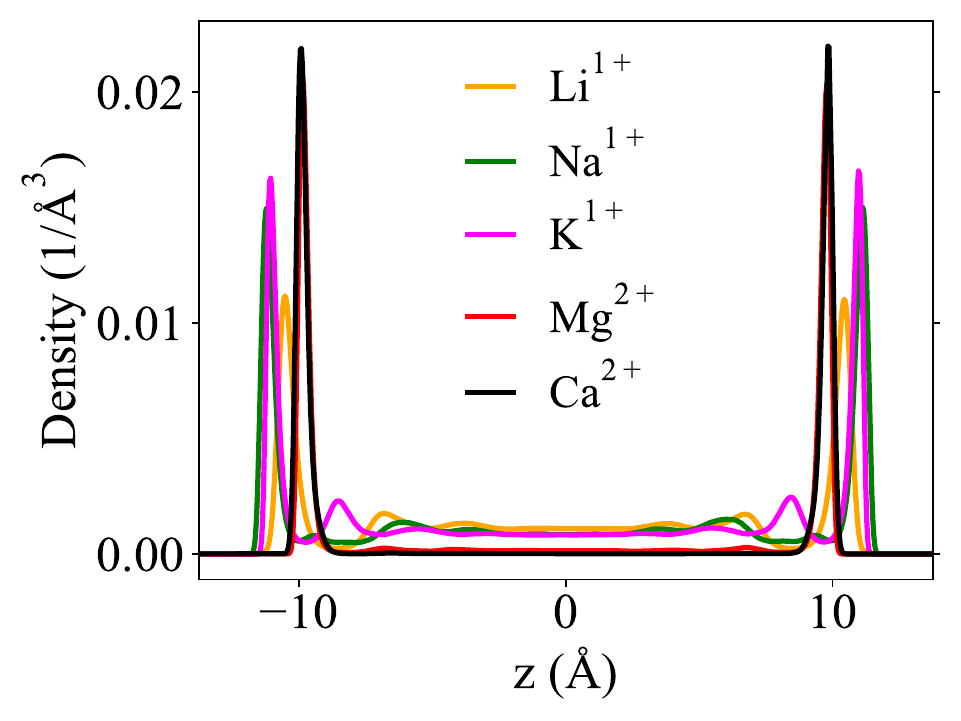}};
        \node[anchor=north west] at ([xshift=-9pt,yshift=5pt]image.north west) {(a)};
        \end{tikzpicture}
    \end{subfigure}
    \hspace{1em}
    \begin{subfigure}[b]{0.3\linewidth}\label{fig:IONS_DM_LO}
        \centering
        \begin{tikzpicture}
        \node[anchor=south west, inner sep=0] (image) at (0,0)
        {\includegraphics[trim={0.0em 0.0em 0.0em 0.0em},clip,height = 4.2cm,width=1.0\textwidth]{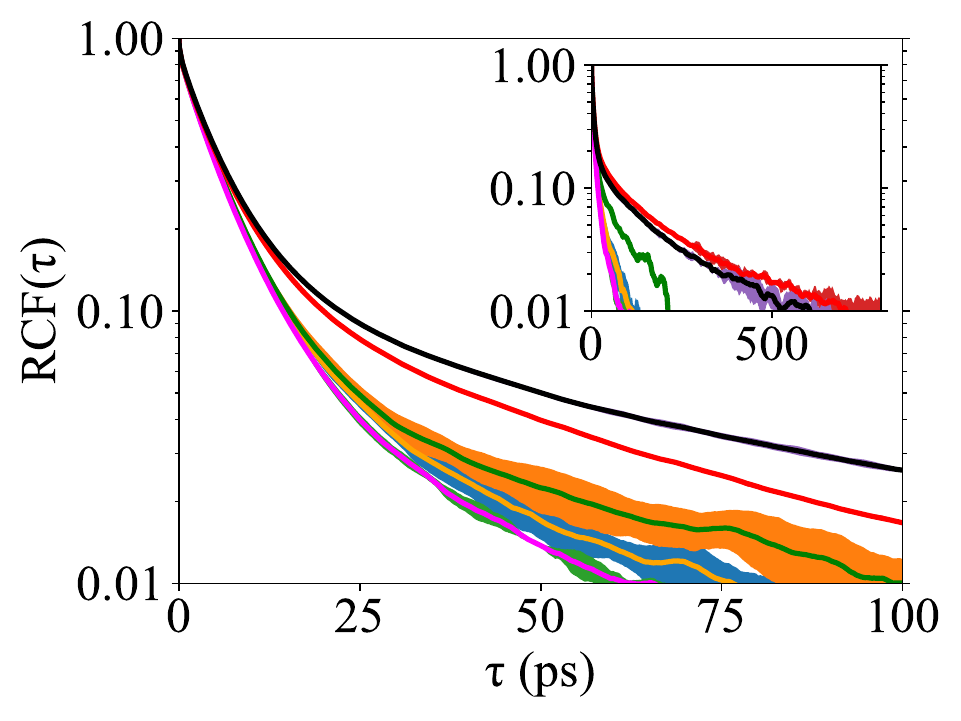}};
        \node[anchor=north west] at ([xshift=-9pt,yshift=5pt]image.north west) {(b)};
        \end{tikzpicture}
    \end{subfigure}  
    \hspace{1em}
    \begin{subfigure}[b]{0.3\linewidth}\label{fig:IONS_DM_HI}
        \centering
        \begin{tikzpicture}
        \node[anchor=south west, inner sep=0] (image) at (0,0)
        {\includegraphics[trim={0.0em 0.0em 0.0em 0.0em},clip,height = 4.2cm,width=1.0\textwidth]{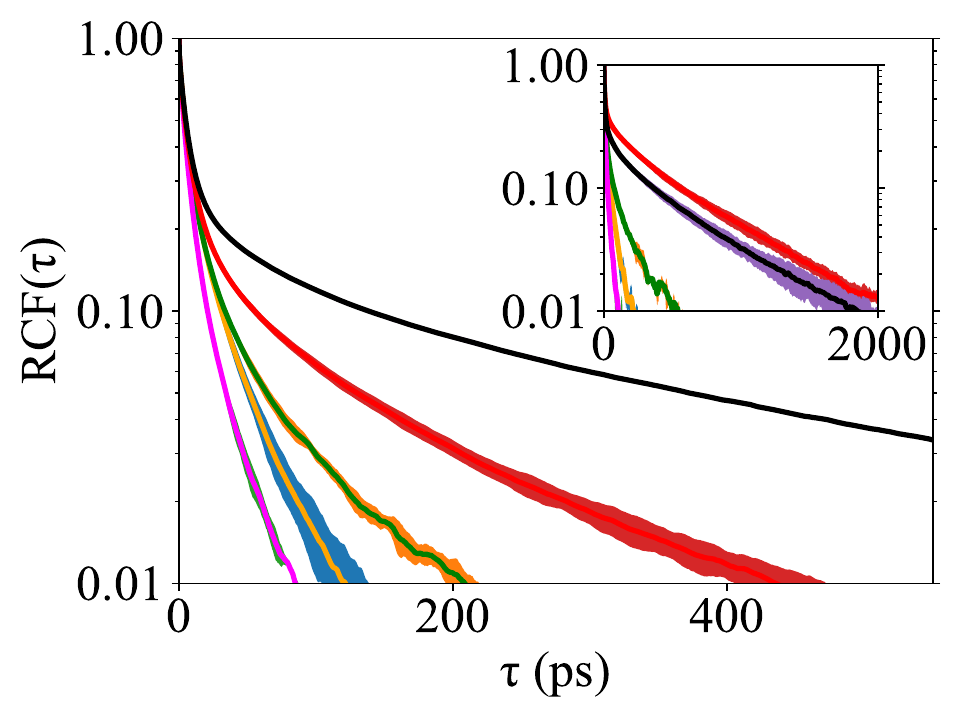}};
        \node[anchor=north west] at ([xshift=-9pt,yshift=5pt]image.north west) {(c)};
        \end{tikzpicture}
    \end{subfigure}  
    \caption{(a) Atomic density profile along the $z$-axis perpendicular to the surface for  high values of the concentration of ions. Reorientation correlation function of water dipole moment at (b) low (0.5M) and (c) high concentrations of ions (1.5M and 2.5M) in the central region of the liquid (R3). Insets show RCF($\tau$) in the interfacial region (R1, R5). Shaded areas represent standard deviations obtained with 3 independent MD runs.} 
    \label{fig:SA_IONS_MD}
\end{figure}

\renewcommand{\thetable}{SV}
\begin{table}[!ht]
    \begin{center}
    \caption{Reorientation times of water dipole moment for all regions of the liquid with low concentration of salt (0.5 M for all cations). Uncertainties in $\tau_{DM}$ are given in the last digits in parenthesis.}
    \centering
    \begin{tabular}{ccccccccccccc}
    \hline
    \hline
    Region & \multicolumn{2}{c}{Water} & \multicolumn{2}{c}{Water/LiCl} & \multicolumn{2}{c}{Water/NaCl} & \multicolumn{2}{c}{Water/KCl} & \multicolumn{2}{c}{Water/MgCl$_2$} & \multicolumn{2}{c}{Water/CaCl$_2$} \\ [3pt]
     & $\tau_{DM}$, ps & $\beta$ & $\tau_{DM}$, ps & $\beta$ & $\tau_{DM}$, ps & $\beta$ & $\tau_{DM}$, ps & $\beta$ & $\tau_{DM}$, ps & $\beta$ & $\tau_{DM}$, ps & $\beta$ \\ [3pt]
    \hline
    \rowcolor{Gray}
    R1 & 4.47(4) & 0.74(1) & 5.31(3) & 0.61(1) & 5.15(6) & 0.54(2) & 4.89(4) & 0.65(1) & 5.5(3) & 0.33(1) & 5.00(5) & 0.55(2) \\
    R2 & 4.30(3) & 0.74(1) & 4.86(2) & 0.64(1) & 4.79(6) & 0.62(1) & 4.69(4) & 0.66(1) & 4.7(1) & 0.31(1) & 4.70(3) & 0.66(1) \\
    \rowcolor{Gray}
    R3 & 4.21(1) & 0.75(1) & 4.71(2) & 0.66(1) & 4.63(1) & 0.64(1) & 4.55(1) & 0.68(1) & 5.14(1) & 0.56(1) & 4.59(1) & 0.69(1) \\
    R4 & 4.30(2) & 0.75(1) & 4.89(4) & 0.63(1) & 4.74(3) & 0.63(1) & 4.70(2) & 0.67(1) & 4.6(1) & 0.30(1) & 4.70(2) & 0.66(1) \\
    \rowcolor{Gray}
    R5 & 4.47(3) & 0.74(1) & 5.29(5) & 0.60(1) & 5.16(9) & 0.52(1) & 4.91(6) & 0.65(1) & 5.5(2) & 0.32(1) & 5.01(5) & 0.54(2) \\
    \hline
    \hline
    \end{tabular}
    \label{table:SA_TAU_DM_LOW}
    \end{center}
\end{table}

\renewcommand{\thetable}{SVI}
\begin{table}[!ht]
    \begin{center}
    \caption{Reorientation times of water dipole moment for all regions of the channel with high concentration of salt (2.5 M, monovalent cations). Uncertainties in $\tau_{DM}$ are given in the last digits in parenthesis.}
    \centering
    \begin{tabular}{ccccccccc}
    \hline
    \hline
    Region & \multicolumn{2}{c}{Water} & \multicolumn{2}{c}{Water/LiCl} & \multicolumn{2}{c}{Water/NaCl} & \multicolumn{2}{c}{Water/KCl} \\ [3pt]
     & $\tau_{DM}$, ps & $\beta$ & $\tau_{DM}$, ps & $\beta$ & $\tau_{DM}$, ps & $\beta$ & $\tau_{DM}$, ps & $\beta$ \\ [3pt]
    \hline
    \rowcolor{Gray}
    R1 & 4.47(4) & 0.74(1) & 8.32(8) & 0.55(1) & 8.44(9) & 0.44(1) & 6.22(5) & 0.60(1) \\
    R2 & 4.30(3) & 0.74(1) & 8.14(7) & 0.55(1) & 7.42(2) & 0.49(1) & 5.95(2) & 0.61(1) \\
    \rowcolor{Gray}
    R3 & 4.21(1) & 0.75(1) & 7.78(2) & 0.56(1) & 7.13(4) & 0.50(1) & 5.87(2) & 0.62(2) \\
    R4 & 4.30(2) & 0.75(1) & 8.12(4) & 0.55(1) & 7.46(9) & 0.48(1) & 5.96(2) & 0.61(1) \\
    \rowcolor{Gray}
    R5 & 4.47(3) & 0.74(1) & 8.35(9) & 0.55(1) & 8.40(8) & 0.44(1) & 6.20(2) & 0.61(1) \\
    \hline
    \hline
    \end{tabular}
    \label{table:SA_TAU_DM_MONO_HIGH}
    \end{center}
\end{table}

\renewcommand{\thetable}{SVII}
\begin{table}[!ht]
    \begin{center}
    \caption{Reorientation times of water dipole moment for all regions of the channel with high concentration of salt (1.5 M, monovalent cations).}
    \centering
    \begin{tabular}{ccccccc}
    \hline
    \hline
    Region & \multicolumn{2}{c}{Water} & \multicolumn{2}{c}{Water/MgCl$_2$} & \multicolumn{2}{c}{Water/CaCl$_2$} \\ [3pt]
     & $\tau_{DM}$, ps & $\beta$ & $\tau_{DM}$, ps & $\beta$ & $\tau_{DM}$, ps & $\beta$ \\ [3pt]
    \hline
    \rowcolor{Gray}
    R1 & 4.47(4) & 0.74(1) & 42.31 $\pm$ 1.93 & 0.36(1) & 17.25 $\pm$ 1.16 & 0.32(1) \\
    R2 & 4.30(3) & 0.74(1) & 36.19 $\pm$ 1.56 & 0.35(1) & 10.81 $\pm$ 0.41 & 0.30(1) \\
    \rowcolor{Gray}
    R3 & 4.21(1) & 0.75(1) & 7.54 $\pm$ 0.1 & 0.39(1) & 8.91 $\pm$ 0.21 & 0.31(1) \\
    R4 & 4.30(2) & 0.75(1) & 35.17 $\pm$ 1.75 & 0.35(1) & 10.8 $\pm$ 0.41 & 0.31(1) \\
    \rowcolor{Gray}
    R5 &4.47(3) & 0.74(1) & 43.03 $\pm$ 2.01 & 0.37(1) & 17.57 $\pm$ 1.49 & 0.32(1) \\
    \hline
    \hline
    \end{tabular}
    \label{table:SA_TAU_DM_DI_HIGH}
    \end{center}
\end{table}

Results of the average number of hydrogen bonds along the $z$-axis perpendicular to the surface are shown in Fig. \ref{fig:NHB_ALL}. We note that $\langle$n$_{HB}$$\rangle$ changes near the graphene surface, relative to the bulk region, for all salts used in this work. This effect is more dramatic in the case of divalent cations.

\renewcommand{\thefigure}{S10}
\begin{figure}[!ht]
    \begin{subfigure}[b]{0.3\linewidth}\label{fig:NHB_LICL}
        \centering
        \begin{tikzpicture}
        \node[anchor=south west, inner sep=0] (image) at (0,0)
        {\includegraphics[trim={0.0em 0.0em 0.0em 0.0em},clip,height = 4.2cm,width=1.0\textwidth]{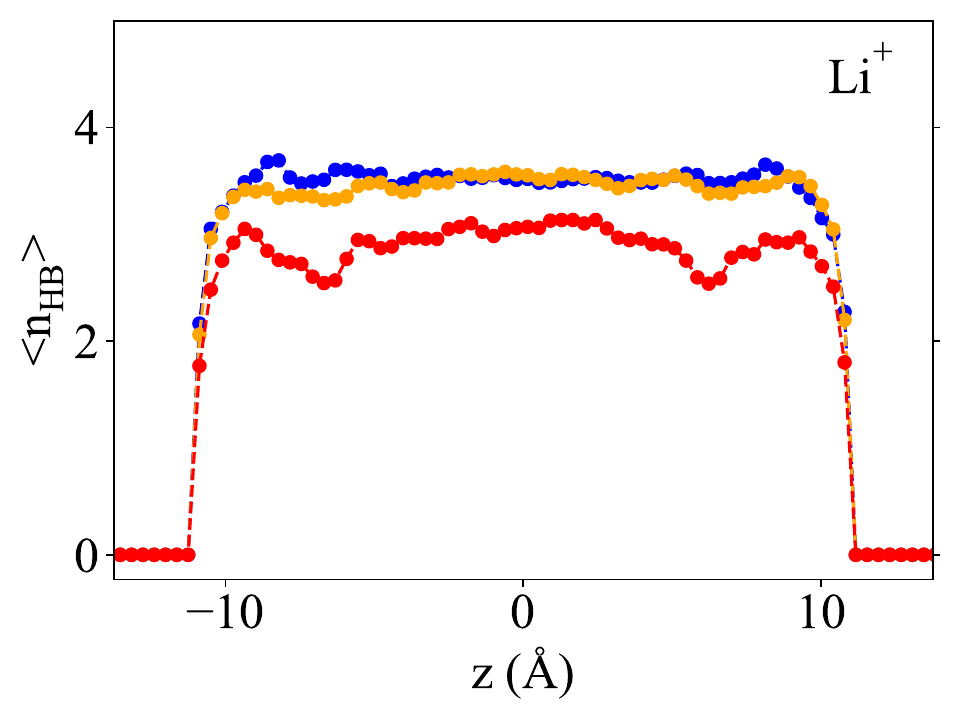}};
        \node[anchor=north west] at ([xshift=-9pt,yshift=5pt]image.north west) {(a)};
        \end{tikzpicture}
    \end{subfigure}
    \hspace{1em}
    \begin{subfigure}[b]{0.3\linewidth}\label{fig:NHB_NACL}
        \centering
        \begin{tikzpicture}
        \node[anchor=south west, inner sep=0] (image) at (0,0)
        {\includegraphics[trim={0.0em 0.0em 0.0em 0.0em},clip,height = 4.2cm,width=1.0\textwidth]{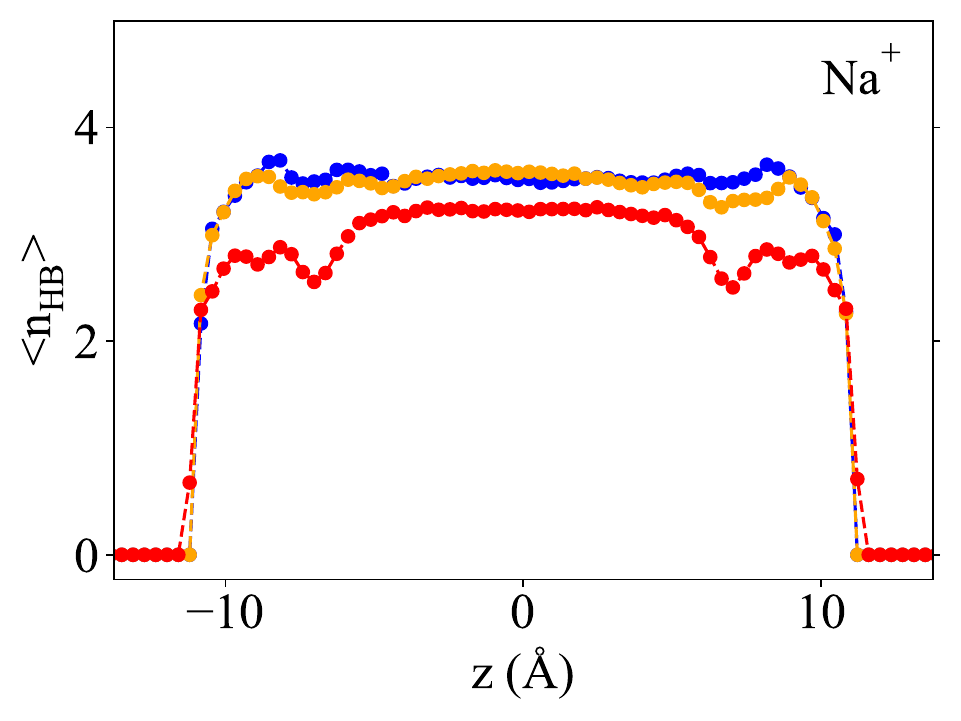}};
        \node[anchor=north west] at ([xshift=-9pt,yshift=5pt]image.north west) {(b)};
        \end{tikzpicture}
    \end{subfigure}
    \hspace{1em}
    \begin{subfigure}[b]{0.3\linewidth}\label{fig:NHB_KCL}
        \centering
        \begin{tikzpicture}
        \node[anchor=south west, inner sep=0] (image) at (0,0)
        {\includegraphics[trim={0.0em 0.0em 0.0em 0.0em},clip,height = 4.2cm,width=1.0\textwidth]{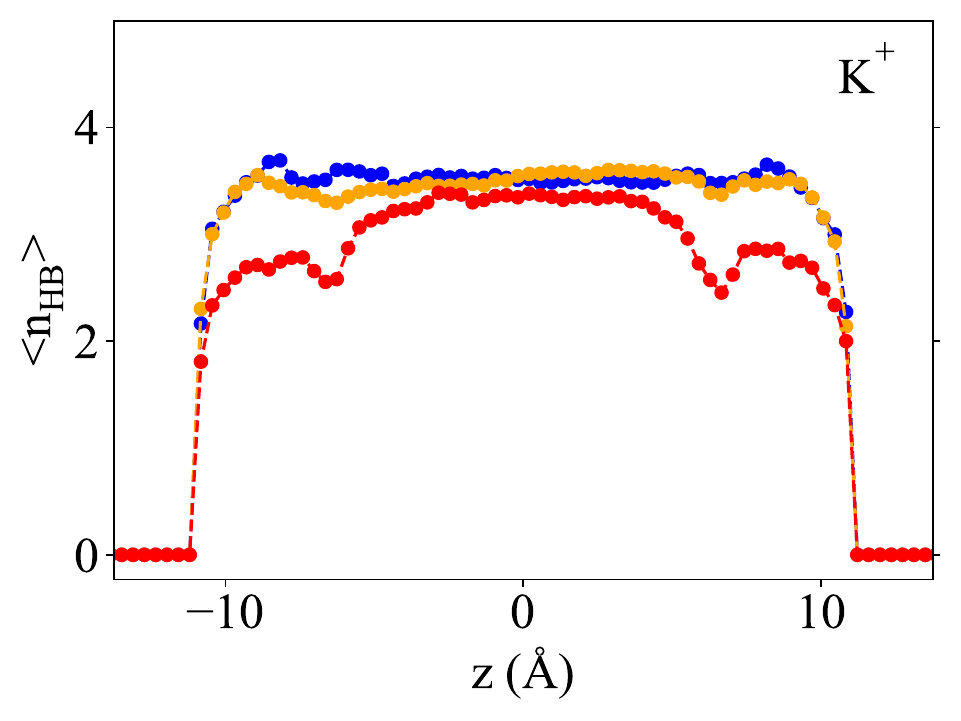}};
        \node[anchor=north west] at ([xshift=-9pt,yshift=5pt]image.north west) {(c)};
        \end{tikzpicture}
    \end{subfigure}
    \hspace{1em}
    \begin{subfigure}[b]{0.3\linewidth}\label{fig:NHB_MGCL}
        \centering
        \begin{tikzpicture}
        \node[anchor=south west, inner sep=0] (image) at (0,0)
        {\includegraphics[trim={0.0em 0.0em 0.0em 0.0em},clip,height = 4.2cm,width=1.0\textwidth]{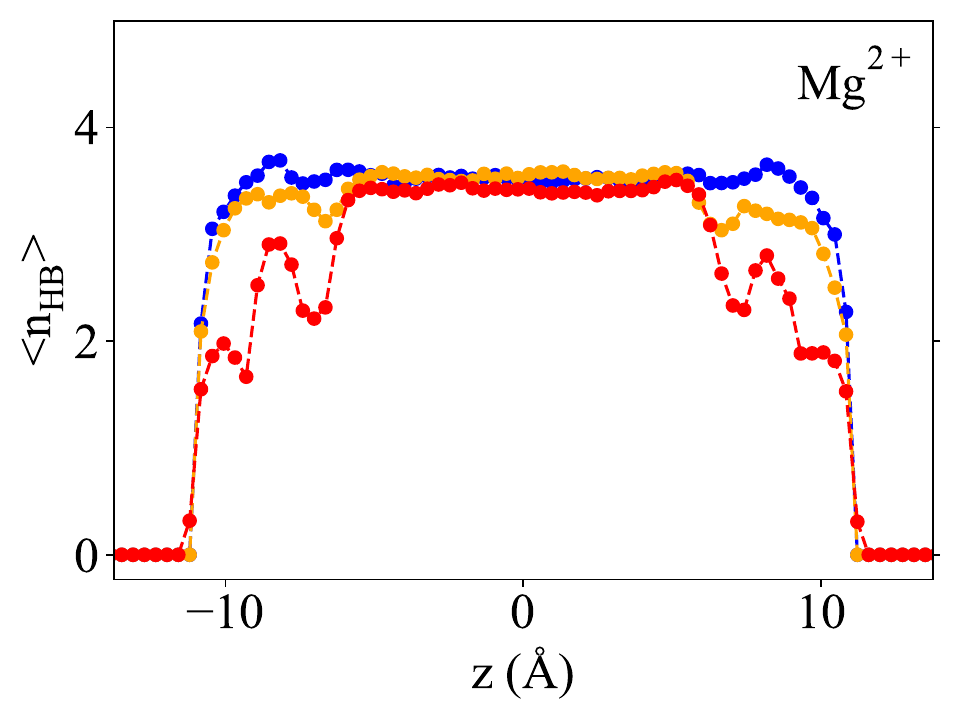}};
        \node[anchor=north west] at ([xshift=-9pt,yshift=5pt]image.north west) {(d)};
        \end{tikzpicture}
    \end{subfigure}  
    \hspace{1em}
    \begin{subfigure}[b]{0.3\linewidth}\label{fig:NHB_CACL}
        \centering
        \begin{tikzpicture}
        \node[anchor=south west, inner sep=0] (image) at (0,0)
        {\includegraphics[trim={0.0em 0.0em 0.0em 0.0em},clip,height = 4.2cm,width=1.0\textwidth]{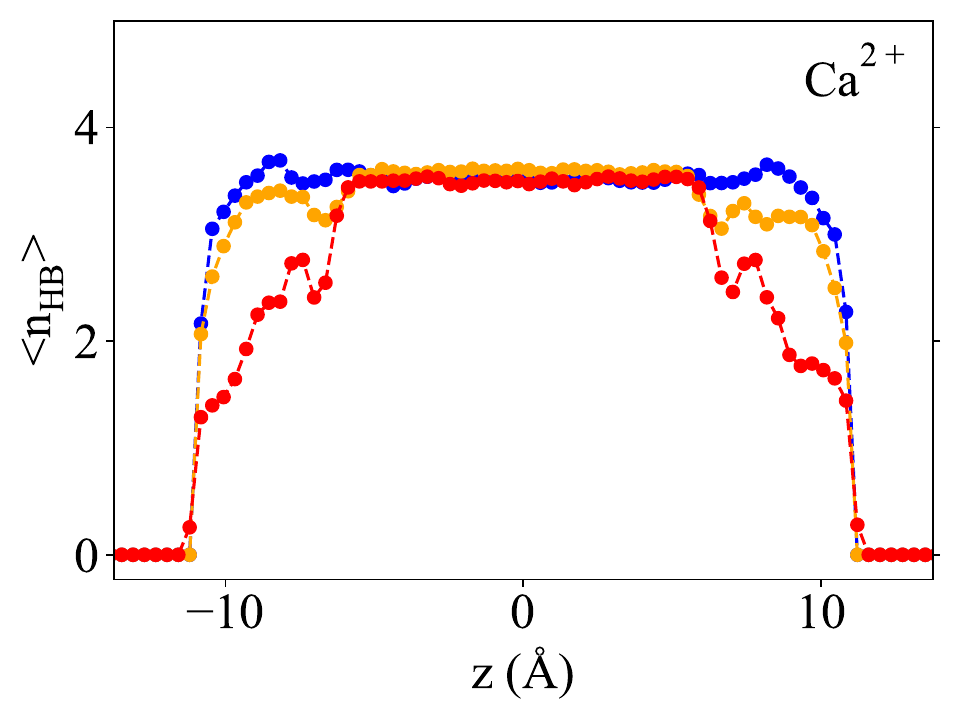}};
        \node[anchor=north west] at ([xshift=-9pt,yshift=5pt]image.north west) {(e)};
        \end{tikzpicture}
    \end{subfigure}
    \caption{Distribution of the average number of hydrogen bonds along the $z$-direction perpendicular to the surface, for all cations and concentrations considered in this work. Values for water, low, and high concentrations of ions are represented by blue, orange, and red colors, respectively.}
    \label{fig:NHB_ALL}
\end{figure}

The results of the average lifetime of hydrogen bonds in all different regions of the supercell with low and high concentrations of ions are shown in Table \ref{table:AVE_LIFE_HBONDS_LOW} and Table \ref{table:AVE_LIFE_HBONDS_HIGH}, respectively.

\renewcommand{\thetable}{SVIII}
\begin{table}[!ht]
    \begin{center}
    \caption{Average lifetime of hydrogen bonds in all regions of the liquid with low concentration of salt (0.5 M). Uncertainties of $\tau_{HB}$ are given in the last digits in parenthesis.}
    \begin{tabular}{ccccccc}
    \hline
    \hline
    Region & Water & Water/LiCl & Water/NaCl & Water/KCl & Water/MgCl$_2$ & Water/CaCl$_2$ \\ [3pt]
     & $\tau_{HB}$, ps & $\tau_{HB}$, ps & $\tau_{HB}$, ps & $\tau_{HB}$, ps & $\tau_{HB}$, ps & $\tau_{HB}$, ps \\ [3pt]
    \hline
    \rowcolor{Gray}
    R1 & 0.69(3) & 0.76(5) & 0.69(2) & 0.76(1) & 0.93(9) & 0.84(3) \\
    R2 & 0.62(1) & 0.72(3) & 0.70(2) & 0.72(3) & 0.97(9) & 0.83(3) \\
    \rowcolor{Gray}
    R3 & 0.70(1) & 0.71(3) & 0.73(1) & 0.72(1) & 0.83(3) & 0.82(4) \\
    R4 & 0.67(1) & 0.70(2) & 0.69(3) & 0.73(3) & 0.95(7) & 0.81(2) \\
    \rowcolor{Gray}
    R5 & 0.68(2) & 0.74(4) & 0.69(3) & 0.74(3) & 0.92(8) & 0.82(2) \\
    \hline
    \hline
    \end{tabular}
    \label{table:AVE_LIFE_HBONDS_LOW}
    \end{center}
\end{table}

\renewcommand{\thetable}{SIX}
\begin{table}[!ht]
    \begin{center}
    \caption{Average lifetime of hydrogen bonds in all regions of the channel with high concentration of salt (1.5 M, and 2.5 M for divalent and monovalent cations, respectively). Uncertainties of $\tau_{HB}$ are given in the last digits in parenthesis.}
    \begin{tabular}{ccccccc}
    \hline
    \hline
    Region & Water & Water/LiCl & Water/NaCl & Water/KCl & Water/MgCl$_2$ & Water/CaCl$_2$ \\ [3pt]
     & $\tau_{HB}$, ps & $\tau_{HB}$, ps & $\tau_{HB}$, ps & $\tau_{HB}$, ps & $\tau_{HB}$, ps & $\tau_{HB}$, ps \\ [3pt]
    \hline
    \rowcolor{Gray}
    R1 & 0.69(3) & 0.72(3) & 0.69(4) & 0.69(4) & 1.4(2) & 0.98(5)  \\
    R2 & 0.62(1) & 0.75(2) & 0.73(5) & 0.73(2) & 1.14(4)  & 0.92(2) \\
    \rowcolor{Gray}
    R3 & 0.70(1) & 0.78(1) & 0.75(1) & 0.70(3) & 0.92(1) & 0.85(2) \\
    R4 & 0.67(1) & 0.75(1)  & 0.73(5) & 0.73(3) & 1.16(2) & 0.89(3) \\
    \rowcolor{Gray}
    R5 & 0.68(2) & 0.73(3) & 0.70(3) & 0.71(5) & 1.3(2) & 0.95(7) \\ 
    \hline
    \hline
    \end{tabular}
    \label{table:AVE_LIFE_HBONDS_HIGH}
    \end{center}
\end{table}

\renewcommand{\thetable}{SX}
\begin{table}[!ht]
    \begin{center}
    \caption{Reorientation times of OH bonds for all regions of the liquid with low concentration of salt (0.5 M, all cations). Uncertainties in $\tau_{DM}$ are given in the last digits in parenthesis.}
    \centering
    \begin{tabular}{ccccccccccccc}
    \hline
    \hline
    Region & \multicolumn{2}{c}{Water} & \multicolumn{2}{c}{Water/LiCl} & \multicolumn{2}{c}{Water/NaCl} & \multicolumn{2}{c}{Water/KCl} & \multicolumn{2}{c}{Water/MgCl$_2$} & \multicolumn{2}{c}{Water/CaCl$_2$} \\ [3pt]
     & $\tau_{OH}$, ps & $\beta$ & $\tau_{OH}$, ps & $\beta$ & $\tau_{OH}$, ps & $\beta$ & $\tau_{OH}$, ps & $\beta$ & $\tau_{OH}$, ps & $\beta$ & $\tau_{OH}$, ps & $\beta$ \\ [3pt]
    \hline
    \rowcolor{Gray}
    R1 & 3.97(2) & 0.79(1) & 4.44(1) & 0.73(1) & 4.43(3) & 0.71(1) & 4.23(3) & 0.75(1) & 5.01(2) & 0.49(3) & 5.00(5) & 0.55(2) \\
    R2 & 3.90(2) & 0.79(1) & 4.25(1) & 0.74(1) & 4.22(3) & 0.74(1) & 4.15(2) & 0.76(1) & 4.84(5) & 0.51(3) & 4.70(3) & 0.66(1) \\
    \rowcolor{Gray}
    R3 & 3.86(1) & 0.79(1) & 4.20(1) & 0.76(1) & 4.16(1) & 0.75(1) & 4.10(1) & 0.77(1) & 4.51(1) & 0.71(1) & 4.59(1) & 0.69(1) \\
    R4 & 3.90(1) & 0.79(1) & 4.27(5) & 0.74(1) & 4.21(3) & 0.75(1) & 4.17(1) & 0.76(1) & 484(2) & 0.51(3) & 4.70(2) & 0.66(1) \\
    \rowcolor{Gray}
    R5 & 3.97(3) & 0.78(1) & 4.40(3) & 0.72(1) & 4.47(4) & 0.70(1) & 4.27(4) & 0.75(1) & 5.02(3) & 0.51(3) & 5.01(5) & 0.53(2) \\
    \hline
    \hline
    \end{tabular}
    \label{table:SA_TAU_OH_LOW}
    \end{center}
\end{table}

\renewcommand{\thetable}{SXI}
\begin{table}[!ht]
    \begin{center}
    \caption{Reorientation times of water dipole moment for all regions of the liquid with high concentration of salt (2.5 M, monovalent cations). Uncertainties in $\tau_{DM}$ are given in the last digits in parenthesis.}
    \centering
    \begin{tabular}{ccccccccc}
    \hline
    \hline
    Region & \multicolumn{2}{c}{Water} & \multicolumn{2}{c}{Water/LiCl} & \multicolumn{2}{c}{Water/NaCl} & \multicolumn{2}{c}{Water/KCl} \\ [3pt]
     & $\tau_{OH}$, ps & $\beta$ & $\tau_{OH}$, ps & $\beta$ & $\tau_{OH}$, ps & $\beta$ & $\tau_{OH}$, ps & $\beta$ \\ [3pt]
    \hline
    \rowcolor{Gray}
    R1 & 3.97(2) & 0.79(1) & 6.07(4) & 0.66(1) & 6.09(4) & 0.58(1) & 5.31(1) & 0.70(1) \\
    R2 & 3.90(2) & 0.79(1) & 5.98(3) & 0.66(1) & 5.76(2) & 0.63(1) & 5.19(1) & 0.71(1) \\
    \rowcolor{Gray}
    R3 & 3.86(1) & 0.79(1) & 5.89(1) & 0.67(1) & 5.64(1) & 0.65(1) & 5.15(1) & 0.71(1) \\
    R4 & 3.90(1) & 0.79(1) & 5.96(2) & 0.66(1) & 5.77(5) & 0.63(1) & 5.20(2) & 0.71(1) \\
    \rowcolor{Gray}
    R5 & 3.97(3) & 0.78(1) & 6.07(3) & 0.66(1) & 6.06(3) & 0.59(1) & 5.30(1) & 0.70(1)\\
    \hline
    \hline
    \end{tabular}
    \label{table:SA_TAU_OH_MONO_HIGH}
    \end{center}
\end{table}

\renewcommand{\thetable}{SXII}
\begin{table}[!ht]
    \begin{center}
    \caption{Reorientation times of water dipole moment for all regions of the liquid with high concentration of salt (1.5 M, divalent cations).}
    \centering
    \begin{tabular}{ccccccc}
    \hline
    \hline
    Region & \multicolumn{2}{c}{Water} & \multicolumn{2}{c}{Water/MgCl$_2$} & \multicolumn{2}{c}{Water/CaCl$_2$} \\ [3pt]
     & $\tau_{OH}$, ps & $\beta$ & $\tau_{OH}$, ps & $\beta$ & $\tau_{OH}$, ps & $\beta$ \\ [3pt]
    \hline
    \rowcolor{Gray}
    R1 & 3.97(2) & 0.79(1) & 9.2(2) & 0.32(1) & 6.2(3) & 0.33(1) \\
    R2 & 3.90(2) & 0.79(1) & 7.8(3) & 0.31(1) & 5.4(1) & 0.36(1) \\
    \rowcolor{Gray}
    R3 & 3.86(1) & 0.79(1) & 6.33(2) & 0.54(1) & 5.58(7) & 0.40(1) \\
    R4 & 3.90(1) & 0.79(1) & 7.7(2) & 0.32(1) & 5.5(1) & 0.36(2) \\
    \rowcolor{Gray}
    R5 &3.97(3) & 0.78(1) & 9.3(2) & 0.33(1) & 6.2(2) & 0.34(1) \\
    \hline
    \hline
    \end{tabular}
    \label{table:SA_TAU_OH_DI_HIGH}
    \end{center}
\end{table}

\renewcommand{\thefigure}{S11}
\begin{figure}[!ht]
    \begin{subfigure}[b]{0.3\linewidth}\label{fig:TAU_VS_T2_DM}
        \centering
        \begin{tikzpicture}
        \node[anchor=south west, inner sep=0] (image) at (0,0)
        {\includegraphics[trim={0.0em 0.0em 0.0em 0.0em},clip,height=4.2cm, width=1.0\textwidth]{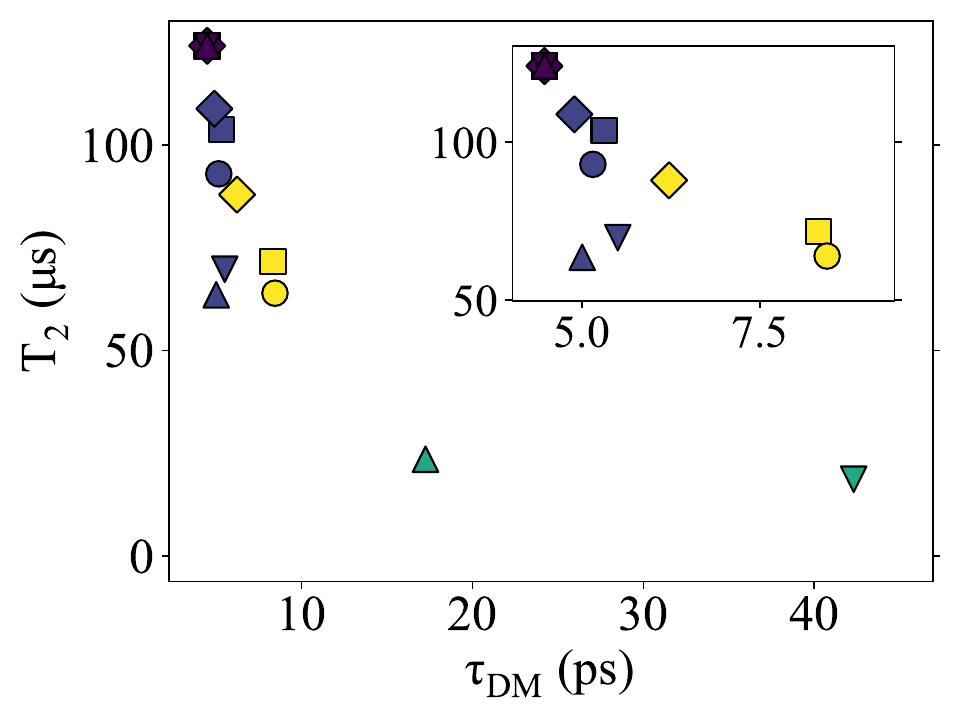}};
        \node[anchor=north west] at ([xshift=-9pt,yshift=5pt]image.north west) {(a)};
        \end{tikzpicture}
    \end{subfigure}    
    \hspace{1em}
    \begin{subfigure}[b]{0.3\linewidth}\label{fig:TAU_VS_T2_HB}
        \centering
        \begin{tikzpicture}
        \node[anchor=south west, inner sep=0] (image) at (0,0)
        {\includegraphics[trim={0.0em 0.0em 0.0em 0.0em},clip,height=4.2cm,width=1.0\textwidth]{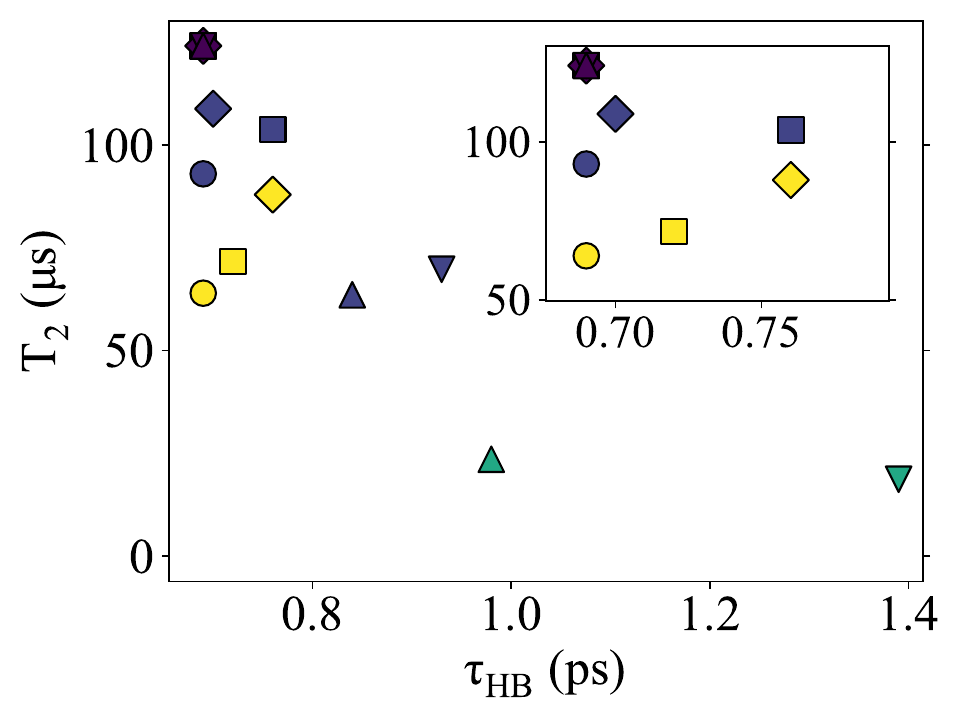}};
        \node[anchor=north west] at ([xshift=-9pt,yshift=5pt]image.north west) {(b)};
        \end{tikzpicture}
    \end{subfigure}  
    \hspace{1em}
    \begin{subfigure}[b]{0.3\linewidth}\label{fig:TAU_VS_T2_OH}
        \centering
        \begin{tikzpicture}
        \node[anchor=south west, inner sep=0] (image) at (0,0)
        {\includegraphics[trim={0.0em 0.0em 0.0em 0.0em},clip,height=4.2cm, width=1.0\textwidth]{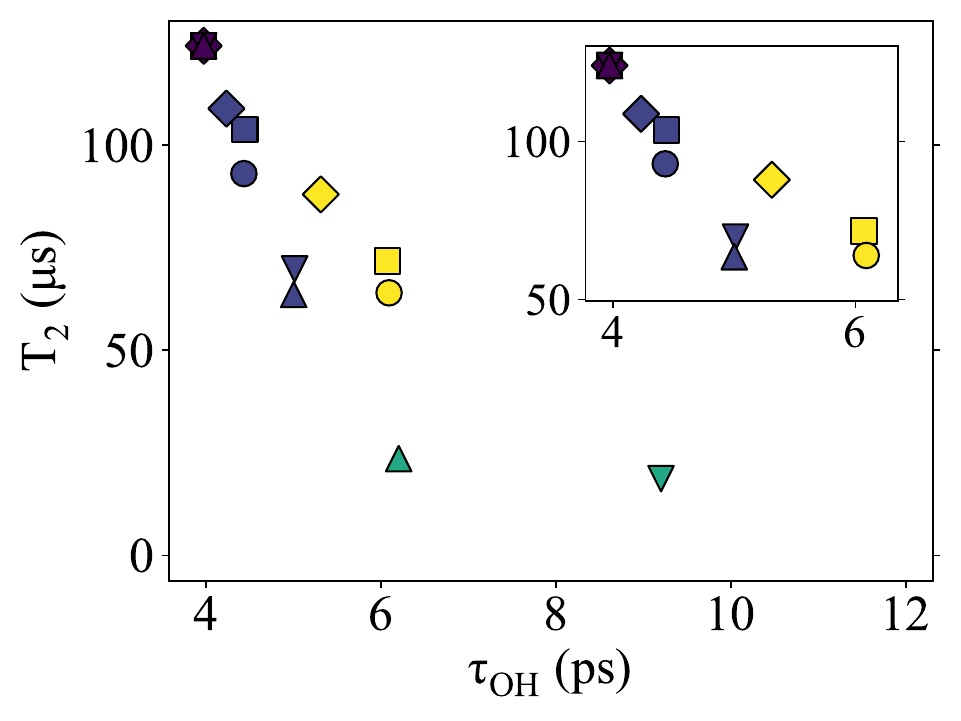}};
        \node[anchor=north west] at ([xshift=-9pt,yshift=5pt]image.north west) {(c)};
        \end{tikzpicture}
    \end{subfigure} 
    \hspace{1em}
    \begin{subfigure}[b]{0.3\linewidth}\label{fig:IONS_labels}
        \centering
        \begin{tikzpicture}
        \node[anchor=south west, inner sep=0] (image) at (0,0)
        {\includegraphics[trim={-6.0em 24.0em 0.0em 0.0em},clip,height=0.8cm, width=1.0\textwidth]{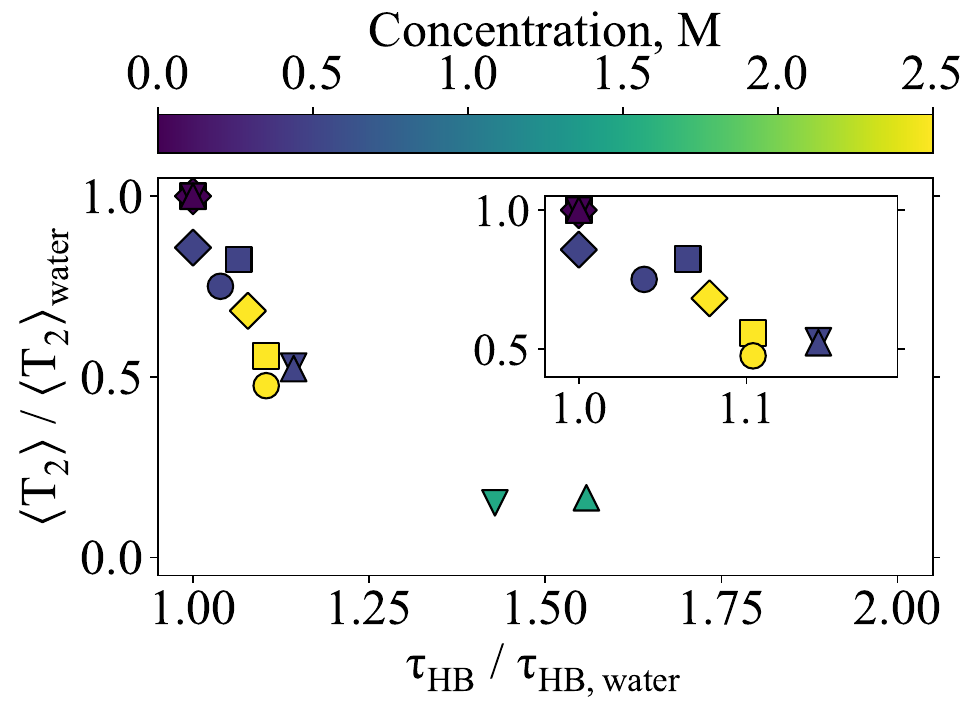}};
        \end{tikzpicture}
    \end{subfigure} 
    \caption{Coherence times T$_2$ as a function of (a) water dipole moment reorientation time, (b) average lifetime of hydrogen bonds, and (c) OH bond reorientation time in the interfacial region R1 of the liquid as a function of concentration for all ions considered in this work: Li$^{+}$ (\protect\markerDone), Na$^{+}$ (\protect\markerDtwo), K$^{+}$ (\protect\markerDthree), Mg$^{2+}$ (\protect\markerDfour), and Ca$^{2+}$ (\protect\markerDfive).}
    \label{fig:TAUS_IONS}
\end{figure}

\renewcommand{\thetable}{SXIII}
\begin{table}
    \begin{center}
    \caption{Fitting parameters of Eq. 7 in the main text that define the correlation functions C($\tau$) for water and ions at low and high concentrations.}
    \begin{tabular}{ccccc}
    \hline
    \hline
    System & Salt Concentration & \multicolumn{3}{c}{Fitting Parameters} \\ [3pt]
     & mol/L & $\Delta$, MHz & n$_c$ & $\tau_c$, ps \\
    \hline
    Water & 0.0 & 5.67 $\pm$ 0.012 & 0.39 & 6.46 $\pm$ 0.16 \\
    \rowcolor{Gray}
    Water/LiCl & 0.5 & 5.74 $\pm$ 0.019 & 0.38 & 7.63 $\pm$ 0.19 \\
     & 2.5 & 5.54 $\pm$ 0.023 & 0.38 & 11.87 $\pm$ 0.31 \\
     \rowcolor{Gray}
    Water/NaCl & 0.5 & 5.72 $\pm$ 0.02 & 0.38 & 8.53 $\pm$ 0.21 \\
     & 2.5 & 5.32 $\pm$ 0.029 & 0.38 & 14.67 $\pm$ 0.37 \\
     \rowcolor{Gray}
     Water/KCl & 0.5 & 5.68 $\pm$ 0.017 & 0.38 & 7.42 $\pm$ 0.19 \\
     & 2.5 & 5.31 $\pm$ 0.024 & 0.38 & 10.7 $\pm$ 0.27 \\
     \rowcolor{Gray}
    Water/MgCl$_2$ & 0.5 & 5.78 $\pm$ 0.032 & 0.37 & 11.33 $\pm$ 0.43 \\
     & 1.5 & 5.69 $\pm$ 0.052 & 0.37 & 42.67 $\pm$ 1.35 \\
     \rowcolor{Gray}
     Water/CaCl$_2$ & 0.5 & 5.79 $\pm$ 0.034 & 0.37 & 12.52 $\pm$ 0.58\\
     & 1.5 & 5.53 $\pm$ 0.045 & 0.35 & 41.45 $\pm$ 1.53 \\
    \hline
    \hline
    \end{tabular}
    \end{center}
\end{table}

\clearpage
\bibliography{References}

\end{document}